\documentclass[%
 aip,
 amsmath,amssymb,
 reprint,%
]{revtex4-1}

\usepackage{graphicx}
\usepackage{dcolumn}
\usepackage{bm}

\usepackage[utf8]{inputenc}
\usepackage[T1]{fontenc}
\usepackage{mathptmx}
\usepackage{etoolbox}

\usepackage{color}

\newcommand{\changed}[1]{{#1}}

\newcommand{\vect}[1]{{\mathbf #1}}
\newcommand{\nue}{\nu_{\mathrm{e}}^*}
\newcommand{\nui}{\nu_{\mathrm{i}}^*}
\newcommand{\nfp}{n_{\mathrm{fp}}}
\newcommand{\etal}{\emph{et al}}

\newcommand{\p}{d}
\newcommand{\mcl}{\ensuremath{\mathcal{L}}}
\newcommand{\dndp}{\ensuremath{\frac{\p \ln{n_{\text{e}}}}{\p \psi_{\text{t}}}}}
\newcommand{\dnidp}{\ensuremath{\frac{\p \ln{n_{\text{i}}}}{\p \psi_{\text{t}}}}}
\newcommand{\dTedp}{\ensuremath{\frac{\p \ln{T_{\text{e}}}}{\p \psi_{\text{t}}}}}
\newcommand{\dTidp}{\ensuremath{\frac{\p \ln{T_{\text{i}}}}{\p \psi_{\text{t}}}}}
\newcommand{\zeff}{\ensuremath{Z_{\text{eff}}}}
\newcommand{\fteff}[1]{\ensuremath{f_{\text{f},#1}^{\text{eff}}}}
\newcommand{\ft}{\ensuremath{f_{\text{t}}}}
\newcommand{\lang}{\left\langle}
\newcommand{\rang}{\right\rangle}

\makeatletter
\def\@email#1#2{%
 \endgroup
 \patchcmd{\titleblock@produce}
  {\frontmatter@RRAPformat}
  {\frontmatter@RRAPformat{\produce@RRAP{*#1\href{mailto:#2}{#2}}}\frontmatter@RRAPformat}
  {}{}
}%
\makeatother
\begin{document}


\title[Optimization of quasisymmetric stellarators with self-consistent bootstrap current and energetic particle confinement]{Optimization of quasisymmetric stellarators with self-consistent bootstrap current and energetic particle confinement}
\author{M. Landreman}
  \email{mattland@umd.edu.}
\author{S. Buller}%
\affiliation{ 
Institute for Research in Electronics and Applied Physics, University of Maryland, College Park MD 20742, USA
}%

\author{M. Drevlak}
\affiliation{%
Max Planck Institute for Plasma Physics, 17491 Greifswald, Germany
}%

\date{\today}

\begin{abstract}
Quasisymmetry can greatly improve the confinement of energetic particles and thermal plasma in a stellarator. The magnetic field of a quasisymmetric stellarator at high plasma pressure is significantly affected by the bootstrap current, but the computational cost of accurate stellarator bootstrap calculations  has precluded use inside optimization.
Here, a new efficient method is demonstrated for optimization of quasisymmetric stellarator configurations such that the bootstrap current profile is consistent with the geometry.
The approach is based on the fact that all neoclassical phenomena in quasisymmetry are isomorphic to those in axisymmetry.
Therefore accurate formulae for the bootstrap current in tokamaks, which can be evaluated rapidly, can be applied also in stellarators. The deviation between this predicted parallel current and the actual parallel current in the magnetohydrodynamic equilibrium is penalized in the objective function, and the current profile of the equilibrium is included in the parameter space. 
Quasisymmetric configurations with significant pressure are thereby obtained with self-consistent bootstrap current and excellent 
confinement.
In a comparison of fusion-produced alpha particle confinement across many stellarators, the new configurations have significantly lower alpha energy losses than many previous designs.

\end{abstract}

\maketitle


\section{Introduction}

Stellarators have many favorable attributes as a fusion reactor concept, including the absence of disruptions, no need for current drive and the associated recirculating power, intrinsic steady-state capability, and the ability to reliably extrapolate to future devices using computational design \citep{boozer21}. Stellarators with quasisymmetry, in which the magnetic field strength has an invariant direction in Boozer coordinates, are of particular interest, as quasisymmetry enables the confinement of fusion-produced alpha particles and reduction of neoclassical transport to acceptable levels \citep{LandremanPaul2022}.
At reactor-relevant densities and temperatures, a quasisymmetric stellarator will have a significant bootstrap current, a current parallel to the magnetic field driven by gradients in the density and temperature. The bootstrap current is a kinetic effect, arising from the different guiding-center trajectories of electrons and ions, as well as collisions. The current in the plasma will produce a contribution to the magnetic field that must be added to the field of the external electromagnetic coils. Accurate reckoning of the bootstrap current is therefore necessary to know the total magnetic field, and hence to know almost any property of interest. 

In this article, a new efficient method is described for including the bootstrap current in the optimization of quasisymmetric stellarators. The approach is a combination of three ideas. The first is a new objective function and associated choice of parameter space for obtaining self-consistency of the current profile. The second is the application of the  isomorphism between quasisymmetric stellarators and tokamaks \citep{Pytte, Boozer83, landreman_isomorphism}, allowing the bootstrap current to be computed as if the geometry was axisymmetric. The third is the application of a recent analytic formula for the tokamak bootstrap current by Redl \etal\cite{Redl}. These last two components greatly reduce the computational cost of stellarator bootstrap current calculations, and as we will show, give quite accurate results compared to bootstrap calculations that fully account for three-dimensional geometry. As examples of the new optimization approach, we demonstrate new quasisymmetric stellarator configurations with self-consistent current profiles, also showing excellent alpha particle confinement and low neoclassical transport.

The bootstrap current on each flux surface depends on the magnetic geometry. A self-consistent state must be determined, a simultaneous solution of the equations of magnetohydrodynamic (MHD) equilibrium and the drift-kinetic equation. 
In the MHD equilibrium problem, a profile of either toroidal current or rotational transform (along with a pressure profile) is provided as an input, and the magnetic field  $\vect{B}$ is an output. Conversely, in solutions of the drift-kinetic equation, the magnetic field components are taken as input, and the parallel current on each flux surface is an output. Generally, if an equilibrium calculation is followed by a drift-kinetic calculation, there is no reason for the resulting parallel current profile to match the current profile of the MHD equilibrium. However, the drift-kinetic profile of parallel current  can be used as input for a new equilibrium calculation, as the basis for a fixed-point iteration. After several iterations between the equilibrium and drift-kinetic calculations, a self-consistent state is obtained \citep{Watanabe92}. While this fixed-point iteration works well for a single configuration, it is not clear how to include this iteration within optimization. If one or a few fixed-point iterations were performed for each evaluation of the objective function, the objective would take on slightly different values each time it is evaluated at the same point in parameter space, potentially confusing the optimization algorithm, and rendering finite difference derivatives very inaccurate. Here we circumvent these issues by using a different approach in which there is no fixed-point iteration. Instead, consistency of the MHD equilibrium and the drift-kinetic equation is enforced through a penalty in the objective.

Another complication of including bootstrap calculations inside stellarator optimization is their computational cost.
Accurate stellarator bootstrap calculations require the solution of the drift-kinetic equation, a high-dimensional 
advection-dominated integro-differential equation, which is numerically challenging. For the case of quasisymmetric stellarators, we can exploit their isomorphism with axisymmetric geometries, identified by Pytte and Boozer\cite{Pytte}. In this isomorphism, formulae for neoclassical phenomena in a quasisymmetric stellarator are the same as the corresponding formulae in axisymmetry, up to a few substitutions. The primary substitution of interest is that the rotational transform $\iota$ in axisymmetry is replaced with $\iota-N$, where $N=0$ for quasi-axisymmetry and $N$ is the number of field periods for quasi-helical symmetry. As a result, the bootstrap current in a quasisymmetric stellarator can be computed as if the geometry were axisymmetric, at lower computational cost due to the ignorable coordinate.

Moreover, for the bootstrap current in axisymmetry, accurate analytic formulae are available, eliminating the need for any numerical kinetic calculation. Here we will use a recent formula for the tokamak bootstrap current  by Redl \etal\cite{Redl},
which improves on an earlier formula by Sauter \etal\cite{sauter}.
The form of this formula is motivated by analytic limits in collisionality and aspect ratio, with coefficients determined by fits to a database of first-principles numerical solutions of the drift-kinetic equation. The resulting formula is applicable at any collisionality and aspect ratio. Here we show for the first time that these analytic tokamak expressions are accurate also in quasisymmetric stellarators.

Expressions for the bootstrap current in a general stellarator are available in the limit of long mean free path compared to system size \citep{ShaingCallen,Shaing89,HelanderBootstrap17}, which can be evaluated faster than the general-collisionality drift-kinetic equation. However, to date these long-mean-free-path expressions tend to produce noisy results \citep{KernbichlerBootstrap}, as they depend on the solutions to magnetic differential equations that are singular on any rational surface. Therefore ad-hoc smoothing of the current profile is sometimes applied. This noisiness is unfavorable for optimization, as smooth objective functions are generally easier to optimize. Also, as will be shown below, the long-mean-free-path formula can give significantly different results in practice to the general-collisionality drift-kinetic equation, the latter being more general and accurate. If improved analytic expressions for the bootstrap current in a general stellarator in the long-mean-free-path regime become available, they could be incorporated into stellarator optimization using the same objective function and parameter space described in this paper.

In the following sections, the quasisymmetry isomorphism is reviewed, and its application to the bootstrap current formula of Redl \etal\cite{Redl} is described. 
In section \ref{sec:one_step}, the new method of calculating the bootstrap current is demonstrated in existing stellarator configurations.
Details of the new optimization procedure are given in section \ref{sec:optimization}.
Optimization results are presented in section \ref{sec:results},
at several values of $\beta = 2 \mu_0 p / B^2$, where $p$ is the plasma pressure and $B=|\vect{B}|$.
Advantages of the isomorphism approach over long-mean-free-path expressions for the stellarator bootstrap current are illustrated in section \ref{sec:bootsj}.
 In section \ref{sec:confinement}, the excellent confinement of these configurations is demonstrated,
and we conclude in section \ref{sec:conclusions}.


\section{Quasisymmetry isomorphism}

In this section we review the isomorphism between quasisymmetric and axisymmetric
configurations \citep{Pytte,Boozer83,landreman_isomorphism}, focusing on the bootstrap current. For this discussion, we consider the magnetic field represented in terms of the Boozer poloidal and toroidal angles $\theta$ and $\varphi$. We introduce a helical angle $\chi=\theta - N \varphi$ for some integer $N$, where $N$ is arbitrary for the moment. 
\changed{
Quasisymmetry can be expressed as the condition  $B=B(\psi,\chi)$, so $B$ depends on $\theta$ and $\varphi$ only through a fixed linear combination of the angles.
We will consider two types of quasisymmetry: quasi-axisymmetry (QA) and quasi-helical (QH) symmetry.
QA is the case in which $N=0$, so $B=B(\psi,\theta)$, while QH symmetry is the case of nonzero $N$, $B=B(\psi,\theta-N\varphi)$. Note that axisymmetry is a special case of QA.
}

\changed{
Now, without assuming quasisymmetry, we proceed to determine the bootstrap current in a general stellarator or tokamak. The calculation begins by writing}
the Boozer coordinate representation of the field,
\begin{align}
    \vect{B} = \nabla\psi\times\nabla\theta + \iota\nabla\varphi\times\nabla\psi
    = \nabla\psi\times\nabla\chi + \tilde\iota\nabla\varphi\times\nabla\psi
\end{align}
and
\begin{eqnarray}
        \label{eq:Boozer}
\vect{B} &=& K(\psi,\theta,\varphi)\nabla\psi + I(\psi)\nabla\theta + G(\psi)\nabla\varphi \\
    &=& K(\psi,\theta,\varphi)\nabla\psi + I(\psi)\nabla\chi + \tilde{G}(\psi)\nabla\varphi. \nonumber
\end{eqnarray}
Here, $\psi$ is the toroidal flux divided by $2\pi$, $\iota$ is the rotational transform, $\tilde\iota=\iota-N$, 
$G(\psi)$ is $\mu_0/(2\pi)$ times the poloidal current outside the flux surface $\psi$, $I(\psi)$ is $\mu_0/(2\pi)$ times the toroidal current inside the flux surface $\psi$,
and $\tilde G(\psi)=G+NI$.
The Jacobian of the $(\psi,\chi,\varphi)$ coordinates is $\sqrt{g}=(\nabla\psi\cdot\nabla\chi\times\nabla\varphi)^{-1}=(\tilde{G}+\tilde\iota I)/B^2=(G+\iota I)/B^2$.

We use the conventional definition of the bootstrap current as $\left< \vect{j}\cdot\vect{B}\right>$, where $\vect{j} = \mu_0^{-1}\nabla\times\vect{B}$ is the current density and $\left< \ldots \right>$ is a flux surface average. 
The flux surface average of any quantity $Q$ is $\left<Q\right>=(1/V')\int_0^{2\pi}d\chi \int_0^{2\pi}d\varphi \, \sqrt{g}Q$
where $V'=\int_0^{2\pi}d\chi \int_0^{2\pi}d\varphi \, \sqrt{g}$.
The bootstrap current can be computed from
\begin{align}
    \left<\vect{j}\cdot\vect{B}\right>
    =\sum_\alpha Z_\alpha e \left< \int d^3v\, v_{||} f_{\alpha} B \right>,
    \label{eq:current_moment}
\end{align}
where $\alpha$ indicates particle species, $Z_\alpha$ is the species charge
in units of the proton charge $e$, $v_{||}$ is the velocity component parallel to $\vect{B}$, and $f_\alpha$ is the distribution function.

We consider the usual case in which the distribution function is approximately a Maxwellian with no mean flow:
$f_\alpha \approx f_{\alpha,0}= n_\alpha [m_\alpha/(2\pi T_\alpha)]^{3/2}\exp\left(-m_\alpha v^2/(2 T_\alpha)\right)$,
where $m_\alpha$, $n_\alpha(\psi)$, and $T_\alpha(\psi)$ are the species mass, density, and temperature, and $v=|\vect{v}|$ is the speed.
Deviations from the Maxwellian are then accounted for in higher order terms in the distribution function. We can divide the correction into a term that is independent of gyrophase, $f_{\alpha,1}$,
and a gyrophase-dependent term with vanishing gyrophase-average, $\tilde{f}_{\alpha,1}$, writing
 $f_\alpha \approx f_{\alpha,0} + f_{\alpha,1} + \tilde{f}_{\alpha,1}$. Since $\tilde{f}_{\alpha,1}$ does not contribute to (\ref{eq:current_moment}), we will not consider this term further. 
 It is convenient to introduce velocity-space variables $\lambda=v_{\perp}^2 / (v^2 B)$ and $\sigma=\mathrm{sign}(v_{||})$ where $v_\perp=\sqrt{v^2-v_{||}^2}$ is the velocity perpendicular to $\vect{B}$,
 so $f_{\alpha,1}=f_{\alpha,1}(\psi,\chi,\varphi,v,\lambda,\sigma)$.
 The distribution $f_{\alpha,1}$ is found by solving the drift-kinetic equation \citep{Hazeltine},
\begin{equation}
v_{||}\nabla_{||} f_{\alpha,1} - \sum_{\gamma} \hat{C}_{\alpha,\gamma}
= -(\vect{v}_{d,\alpha}\cdot\nabla\psi) 
\frac{\partial f_{\alpha,0}}{\partial\psi}.
\label{eq:DKE}
\end{equation}
Here, $\nabla_{||} = \vect{b}\cdot\nabla$, where $\vect{b} = B^{-1}\vect{B}$, the gradient is performed at fixed $v$ and $\lambda$, and the radial guiding center drift is
\begin{align}
    \vect{v}_{d,\alpha}\cdot\nabla\psi = \frac{m_\alpha}{Z_\alpha e B^3} \left( v_{||}^2 + \frac{v_\perp^2}{2}\right)\vect{B}\times\nabla B\cdot\nabla\psi.
\end{align}
(Note that for this radial component, no low-$\beta$ approximation is required to combine the grad-$B$ and curvature drifts.) In (\ref{eq:DKE}), $\hat{C}_{\alpha,\gamma} = C_{\alpha,\gamma}(f_{\alpha,1},f_{\gamma,0}) + C_{\alpha,\gamma}(f_{\alpha,0},f_{\gamma,1})$ is the linearized collision operator, with $C_{\alpha,\gamma}$ the nonlinear Fokker-Planck collision operator \citep{Rosenbluth, HelanderSigmar}.
Once the drift-kinetic equation is solved, the bootstrap current can be computed from (\ref{eq:current_moment}) in the form
\begin{align}
    \left<\vect{j}\cdot\vect{B}\right>
    =\pi\sum_\alpha Z_\alpha e \left< \sum_\sigma \sigma \int_0^{1/B}d\lambda 
\int_0^{\infty}dv \, v^3 B^2     
 f_{\alpha,1} \right>.
    \label{eq:current_moment2}
\end{align}

\changed{
We now introduce the assumption of quasisymmetry:
$B=B(\psi,\chi)$, so $(\partial B / \partial \varphi)_\chi = 0$.}
In this case, the drift-kinetic equation (\ref{eq:DKE}) can be written
\begin{equation}
\frac{v_{||} \tilde\iota B}{\tilde{G}+\tilde\iota I} \frac{\partial f_{\alpha,1}}{\partial \chi} - \sum_{\gamma} \hat{C}_{\alpha,\gamma}
= 
\frac{\partial f_{\alpha,0}}{\partial\psi}
\frac{m_\alpha}{Z_\alpha e B^3}
\left( v_{||}^2 + \frac{v_\perp^2}{2}\right)
\frac{\tilde{G} B^2}{\tilde{G}+\tilde\iota I} \frac{\partial B}{\partial\chi}.
\label{eq:DKE_QS}
\end{equation}
The linearized collision operator can be written in terms of the velocity coordinates $v$ and $v_{||}=\sigma v\sqrt{1-\lambda B}$, and so it introduces dependence on position only through $B$. Similarly, (\ref{eq:DKE_QS}) depends on position only through $\psi$ and $B$, and hence only through $\psi$ and $\chi$.
It can be seen then that the entire drift-kinetic equation has position dependence only through $(\psi,\chi)$, and so its solution $f_{\alpha,1}$
will have this same property. When the  moment (\ref{eq:current_moment2}) of the distribution function is formed, position dependence appears only through $B$ and $\sqrt{g}$, which depend on position only through $(\psi,\chi)$, with no $\varphi$-dependence introduced.
Thus, the equations for computing the bootstrap current in a quasisymmetric stellarator, (\ref{eq:DKE_QS}) and (\ref{eq:current_moment2}), are identical to the equations one can use to compute the bootstrap current in axisymmetry; in both cases only two spatial dimensions appear in the equations.

Due to these observations, we reach the following conclusion. The bootstrap current on a quasisymmetric flux surface of a given nonaxisymmetric field is identical to the bootstrap current on an equivalent axisymmetric flux surface. In this case, ``equivalent'' means that the quantities appearing in (\ref{eq:DKE_QS}) and (\ref{eq:current_moment2}) are matched on that flux surface: $B$, $\tilde\iota$, $\tilde{G}$, $I$, $n_\alpha$, $T_\alpha$, $dn_\alpha/d\psi$, and $dT_{\alpha}/d\psi$. (The equivalent axisymmetric surface need not correspond to any global MHD equilibrium.) Note that $N=0$ in the equivalent axisymmetric configuration, even if $N\ne 0$ in the quasisymmetric configuration. Thus, we arrive at the following rule: to compute the bootstrap current on a quasisymmetric flux surface with parameters $B$, $\iota$, $G$, and $I$, we can compute instead the bootstrap current in an axisymmetric surface with corresponding parameters $B$, $\iota-N$, $G+NI$, and $I$, also matching $n_\alpha$, $T_\alpha$, $dn_\alpha/d\psi$, and $dT_{\alpha}/d\psi$.


\section{Bootstrap current formula of Redl et al.}
\label{sec:redl}

We will use the recent formula for the tokamak bootstrap current by Redl \etal \cite{Redl}. This formula is based on fitting to a large database of first-principles drift-kinetic calculations generated by the code NEO \citep{Belli1} (which should not be confused with the stellarator neoclassical code of the same name.) The formula is applicable across all \changed{tokamak} regimes of collisionality and \changed{for any} aspect ratio.

The expressions for the bootstrap current by Redl \etal~are lengthy and so are relegated to appendix \ref{sec:redlAppendix}. For application in quasisymmetric stellarators, only a few minor modifications are made, detailed in the remainder of this section. 

First, Redl's eq (2) contains the flux function  $I_R=R B_\phi$, where $R$ is the major radius and $B_\phi$ is the toroidal field. (Here we have added the $R$ subscript to $I_R$ to distinguish it from $I$ in (\ref{eq:Boozer}).) In axisymmetry, $I_R$ coincides with $G$. Moreover, the $I_R$ factor in Redl's eq (2) originates from the fact that the inhomogeneous drive term in the drift-kinetic equation, the right-hand side of (\ref{eq:DKE_QS}), is proportional to $\tilde{G}$, which equals $I_R=G$ in axisymmetry. Thus, it is appropriate to replace the $I_R$ factor in \cite{Redl} with $\tilde G$.

Next, in the tokamak expressions of Redl \etal\cite{Redl}, the collisionalities
\begin{eqnarray}
\label{eq:collisionality_tokamak}
\nue &=& 6.921\times 10^{-18} \frac{q R n_e Z \ln\Lambda_e}{T_e^{3/2} \epsilon^{3/2}}, \\
\nui &=& 4.90\times 10^{-18} \frac{q R n_i Z^4 \ln\Lambda_{ii}}{T_i^{3/2} \epsilon^{3/2}}, \nonumber
\end{eqnarray}
appear. Here, $q=1/\iota$ is the safety factor, $R$ is the major radius in meters, the electron density $n_e$ and ion density $n_i$ are measured in meters$^{-3}$, and the electron temperature $T_e$ and ion temperature $T_i$ are measured in eV. In Ref.~\onlinecite{Redl}, the major radius $R(\psi)$ of a given flux surface was defined by $(R_{\max} + R_{\min})/2$, where $R_{\max}(\psi)$ and $R_{\min}(\psi)$ are the maximum and minimum major radius of the surface. The inverse aspect ratio $\epsilon$ was computed as $(R_{\max} - R_{\min})/(2R)$.
We must consider how to evaluate the factors $q R / \epsilon^{3/2}$ in a stellarator.

To do so, we consider the physical origin of the expressions (\ref{eq:collisionality_tokamak}).
The expressions
(\ref{eq:collisionality_tokamak})
arise from the ratio between the collision operator and the parallel streaming term in the drift-kinetic equation, for the trapped part of phase space. The factor $qR/\epsilon^{3/2}$ arises as a product of factors $1/\epsilon$ and $qR/\sqrt{\epsilon}$. The factor $1/\epsilon$ comes from the pitch-angle scattering term in the collision operator: each of the two pitch-angle derivatives is magnified by a factor $1/\sqrt{\epsilon}$ for trapped particles, since trapped particles have a parallel velocity $v_{||} \sim \sqrt{\epsilon} v$, and so they only need to diffuse by an angle $\sim \sqrt{\epsilon}$ to become untrapped. The factor $qR/\sqrt{\epsilon}$ comes from the inverse of  $v_{||} \nabla_{||} \theta$
in the parallel streaming term, again noting $v_{||} \sim \sqrt{\epsilon} v$. In both the collision and streaming terms, the $\epsilon$ factors originate from the size of the trapped region of velocity space, which is determined by $B_{\max}$ and $B_{\min}$. In tokamaks, $R\sim 1/B$ to a good approximation, while in stellarators it is more accurate to evaluate the size of the trapped region in velocity space using $B$ instead of $R$.
Therefore a reasonable generalization of $\epsilon$ in the Redl formula is
\begin{equation}
    \epsilon(\psi) = \frac{B_{\max} - B_{\min}}{B_{\max} + B_{\min}},
\end{equation}
where $B_{\max}(\psi)$ and $B_{\min}(\psi)$ are the maximum and minimum $B$ on the flux surface. This definition reduces to that of Redl \etal~when $R \propto 1/B$, as is approximately true in axisymmetry, but is a better reflection of the size of the trapped region of velocity space in a quasisymmetric stellarator.

The $qR$ factors in (\ref{eq:collisionality_tokamak}) originate from the inverse of $\nabla_{||}\theta$ in the streaming term of the drift-kinetic equation. In a quasisymmetric stellarator, the analogous expression is $1/(\vect{b}\cdot\nabla\chi) = (G + \iota I) / [(\iota-N)B]$. Some kind of average must be performed so that the overall expression is a flux function. We choose
to evaluate (\ref{eq:collisionality_tokamak}) using
\begin{equation}
    qR \to \frac{G + \iota I}{\iota-N} \left< \frac{1}{B}\right>.
\end{equation}
Other choices of the average could be reasonable, and result in only extremely minor differences.

The magnetic geometry also enters Redl's expressions through the effective fraction of trapped particles,
\begin{equation}
f_t = 1 - \frac{3}{4} \left<B^2\right> \int_0^{1/B_{\max}} \frac{\lambda \, d\lambda}{\left<\sqrt{1-\lambda B}\right>},
    \label{eq:ft}
\end{equation}
\changed{a quantity that arises repeatedly in analytic neoclassical calculations for the tokamak banana regime \cite{HelanderSigmar} or stellarator $1/\nu$ regime \cite{ShaingCallen}.}
This expression requires no modification in a stellarator.

Finally, the density and temperature gradients appear in Ref.~\onlinecite{Redl} as $dn_\alpha/d\psi_p$ and $dT_{\alpha}/d\psi_p$ where $2 \pi \psi_p$ is the poloidal flux, and $d\psi_p = \iota d\psi$. Following the isomorphism, we make the substitutions
$dn_\alpha/d\psi_p \to (\iota-N)^{-1} dn_\alpha/d\psi$ and $dT_\alpha/d\psi_p \to (\iota-N)^{-1} dT_\alpha/d\psi$.

When the field strength is not perfectly quasisymmetric, there are several reasonable options available for evaluating $f_t$ and $\epsilon$.
One option is to evaluate (\ref{eq:ft}) directly, including integration over the toroidal angle in the flux surface averages, and use the actual maximum and minimum $B$ to evaluate $\epsilon$.
A second option is to construct Boozer coordinates, remove any Fourier modes of $B(\theta,\varphi)$ that break the symmetry, and evaluate $f_t$ and $\epsilon$ for the resulting perfectly symmetric $B$ pattern. The second option has some extra complexity due to the need to construct Boozer coordinates.
However it is also less sensitive to isolated maxima and minima of $B$ on a flux surface when symmetry is imperfect.
Both methods have been compared in the optimization method described in the next section, and both are effective. Based on experience so far, the Boozer-coordinate approach results in slightly smaller values of the objective function, and so is used for the results that follow.

\changed{
Unfortunately, the Redl formula is not similarly useful for quasi-poloidally symmetric or quasi-isodynamic stellarators. If these optimizations are achieved perfectly, the quasisymmetry isomorphism and other analytic calculations \cite{HelanderNuhrenberg, LandremanCatto} predict $\left<\vect{j}\cdot\vect{B}\right> \propto I(\psi)$, and so eq (C3) in Ref.~\onlinecite{LandremanCatto} gives $\left<\vect{j}\cdot\vect{B}\right>=0$. In practice, the properties of quasi-isodynamic or
quasi-poloidal symmetry are only achieved imperfectly, and one would prefer to have an estimate of the small nonzero values of 
$\left<\vect{j}\cdot\vect{B}\right>$.
Obtaining a fast and smooth calculation of the bootstrap current in imperfectly quasi-poloidally symmetric or quasi-isodynamic stellarators is an important question for future research.
}


\section{Bootstrap current calculations in stellarators}
\label{sec:one_step}

Before embarking on new optimizations, one should verify that the Redl formula with the modifications of section \ref{sec:redl} is indeed accurate in stellarators. We now demonstrate this correspondence, using the quasi-axisymmetric (QA) and quasi-helically symmetric (QH) configurations from Ref.~\onlinecite{LandremanPaul2022}.
For this initial test, no attempt will be made to obtain self-consistency between MHD equilibrium and the drift-kinetic equation. Instead, the bootstrap current will be computed in fixed magnetic fields. This calculation can be thought of as one step of the fixed point iteration described in the introduction. 

\begin{figure*}
  \centering
  \includegraphics[width=7in]{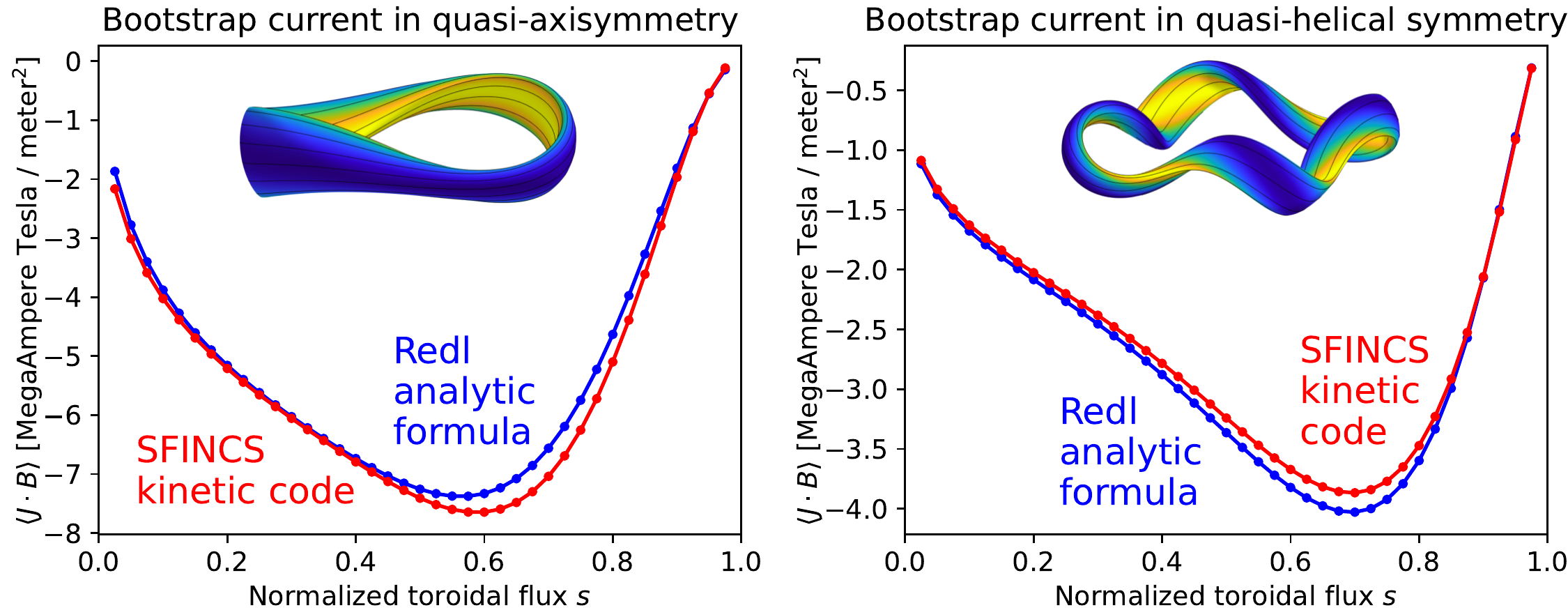}
  \caption{\label{fig:one_step} 
Demonstration that the tokamak bootstrap current formula of Redl \etal\cite{Redl} is a reasonable approximation of full drift-kinetic calculations in quasi-axisymmetric and quasi-helically symmetric stellarators. The two magnetic configurations are from Ref.~\onlinecite{LandremanPaul2022}.
}
\end{figure*}

Both the QA and QH configurations are scaled to 
a minor radius of 1.70 meters and volume-averaged $B$ of 5.86 Tesla, matching the corresponding values of the ARIES-CS reactor study \citep{ARIESCS}.
Here and throughout this paper, density and temperature profiles are specified using
\begin{equation}
n_\alpha(s) = n_{\alpha,0} (1-s^5),
\hspace{0.5in}
T_\alpha(s) = T_{\alpha,0} (1-s),
\label{eq:profiles}
\end{equation}
where $n_{\alpha,0}$ and $T_{\alpha,0}$ are the on-axis values, and $s \in [0,1]$ is the toroidal flux normalized by its value at the plasma boundary. In these first calculations we consider a hydrogen plasma with $n_{e,0}=n_{H,0}=4.13\times 10^{20}$ m$^{-3}$,
and $T_{e,0} = T_{H,0} = 12$ keV. 
(Numerical calculations of the bootstrap current in a hydrogen plasma are nearly indistinguishable from calculations with a deuterium-tritium mixture.)
Here and throughout the remainder of the paper, we assume there are no impurities, so the effective ion charge is $Z_{eff}=1$, although the bootstrap current formula by Redl \etal~can in fact account for impurities.
The results of the modified Redl formula for the two configurations are shown in figure \ref{fig:one_step}.

To assess the accuracy of the Redl formula, the bootstrap current is also calculated using the code SFINCS \citep{sfincs}. This code solves the drift-kinetic equation (\ref{eq:DKE}) in a stellarator, making no assumption of quasisymmetry, and therefore fully accounting for three-dimensional geometry and imperfections in the quasisymmetry. The bootstrap current computed by SFINCS using (\ref{eq:current_moment2}) is displayed in figure \ref{fig:one_step}.

The drift-kinetic equation solved by SFINCS can also include the effect of a radial electric field through the term
$\vect{v}_E \cdot\nabla f_{\alpha,1}$, where
$\vect{v}_E=(d\Phi/d\psi)B^{-2}\vect{B}\times\nabla\psi$ is the $\vect{E}\times\vect{B}$ drift and  $\Phi(\psi)$ is the electrostatic potential.
However, in the SFINCS calculations here, it is very difficult to accurately determine an ambipolar radial electric field, since for any value of the electric field the radial current is found to be within discretization error of zero. This ``intrinsic ambipolarity'' is a consequence of the precise quasisymmetry of these configurations \citep{HelanderSimakov}. Another consequence is that the bootstrap current computed by SFINCS is  independent of the radial electric field to high precision. Therefore, for all SFINCS calculations in this paper, we set the radial electric field to zero.

It is apparent from figure \ref{fig:one_step} that the generalized Redl formula is reasonably accurate when compared to the first-principles SFINCS calculations. 
The small disagreement can arise \changed{in principle} both due to imperfections in the quasisymmetry and due to the fact that even in axisymmetry, the Redl formula is only an approximate fit to solutions of the drift-kinetic equation.
\changed{Additional SFINCS calculations for the two configurations were carried out based on a Boozer-coordinate representation for the magnetic field, filtering out all Fourier modes of $B$ that break quasisymmetry. The results are nearly indistinguishable from the SFINCS calculations that include the imperfections in the quasisymmetry. Therefore for these geometries, the disagreement in figure \ref{fig:one_step} is dominated by the fact that the Redl formula only approximates true drift-kinetic solutions.}
This small disagreement is acceptable for the purposes of optimization, and it is within the large uncertainty associated with predicting the pressure profile. Moreover, as will be shown later, this small difference in the bootstrap current can be eliminated by applying a few steps of fixed-point iteration at the end of an optimization. Given the success of this initial test of the Redl formula in stellarators, we now proceed to use it in the optimization of new stellarator configurations.


\section{Optimization problem}
\label{sec:optimization}

To optimize for quasi-helically symmetric equilibria, we minimize the objective function
\begin{equation}
    f_{QH} = f_{QS} + f_{boot} + (A - A_*)^2 + (a - a_*)^2 + (\bar{B} - \bar{B}_*)^2
    \label{eq:fQH}
\end{equation}
where $A$ is the aspect ratio, $a$ is the minor radius, $\bar{B}$ is the volume-averaged field strength, and quantities with a subscript $*$ are target values. The term $f_{QS}$ represents the departure from quasisymmetry and is
\begin{equation}
\label{eq:fqs}
f_{QS} = \sum_{s_j} w_j \left<
\left( \frac{1}{B^3} \left[
(N - \iota )\vect{B}\times\nabla B \cdot\nabla\psi
-(G+NI) \vect{B}\cdot\nabla B\right]\right)^2
\right>.
\end{equation}
The term $f_{QS}$ or similar expressions have been used previously to optimize for quasisymmetry \cite{Rodriguez,LandremanPaul2022,Dudt}. Finally, the term $f_{boot}$ penalizes inconsistency in the bootstrap current:
\begin{equation}
    f_{boot} = \frac{\int_0^1 ds \left[ \left<\vect{j}\cdot\vect{B}\right>_{\mathrm{vmec}}
    -\left<\vect{j}\cdot\vect{B}\right>_{\mathrm{Redl}}\right]^2}
    {\int_0^1 ds \left[ \left<\vect{j}\cdot\vect{B}\right>_{\mathrm{vmec}}
    +\left<\vect{j}\cdot\vect{B}\right>_{\mathrm{Redl}}\right]^2}.
    \label{eq:fboot}
\end{equation}
Thus, $f_{boot}=0$ if the bootstrap current is self-consistent, whereas if the parallel current in the MHD equilibrium is zero (as it may be in the first iteration), the normalizing denominator in (\ref{eq:fboot}) results in $f_{boot}=1$.
Overall, the objective (\ref{eq:fQH}) promotes quasisymmetry and bootstrap current self-consistency, while keeping the aspect ratio, device size, and average field strength close to desired target values. For all results shown here, we match the minor radius and field strength of ARIES-CS, $a_*=1.70$ meters and $\bar{B}_*=5.86$ Tesla. \changed{We did not find it necessary to introduce weights in front of each term in \ref{eq:fQH}.}

For quasi-axisymmetry, one more term is included in the objective:
\begin{equation}
    f_{QA} = f_{QH} + (\bar{\iota} - \bar{\iota}_*)^2,
    \label{eq:QA_mean_iota_target}
\end{equation}
where $\bar{\iota}=\int_0^1 ds\, \iota$ is the average rotational transform, and $\bar{\iota}_*$ is a target value. The additional term in quasi-axisymmetry is required to prevent the optimum from being axisymmetric\changed{, i.e. a tokamak}.

The integrals in the objective are discretized as finite sums, yielding a nonlinear least-squares problem. In the quasisymmetry term, the flux surface averages are discretized using uniform grids in
the standard toroidal angle $\phi$ 
and in a poloidal angle $\vartheta$, which need not be the Boozer angle.
We can then write 
\begin{equation}
    f_{QS} = \sum_{s_j, \vartheta_k, \phi_\ell} R_{QS}(s_j, \vartheta_k, \phi_\ell)^2,
\end{equation}
where
\begin{eqnarray}
    R_{QS} &=& \sqrt{w_j \frac{ \Delta_{\vartheta} \Delta_\phi}{V_s'}|\sqrt{g_s}|}
    \frac{1}{B^3}
    \\
    &&\times
    \left[(N-\iota )\vect{B}\times\nabla B \cdot\nabla\psi
    -(G+NI) \vect{B}\cdot\nabla B\right],
    \nonumber
\end{eqnarray}
with $\Delta_{\vartheta}$ and $\Delta_\phi$ denoting the grid spacing, 
 $\sqrt{g_s}=1/(\nabla s \cdot \nabla\vartheta\times\nabla\phi)$ is the Jacobian
of the $\{s, \vartheta, \phi\}$ coordinates, and
$V_s'=\Delta_{\vartheta} \Delta_\phi\sum_{\vartheta_k, \phi_\ell}\sqrt{g_s}$.

The parameter space for the optimization consists of the shape of the toroidal boundary surface, the enclosed toroidal flux, and a profile of either toroidal current or $\iota(s)$. The boundary shape is represented as
\begin{eqnarray}
R(\vartheta,\phi) &=& \sum_{m,n}R_{m,n}\cos(m\vartheta-\nfp n\phi),
\\
Z(\vartheta,\phi) &=& \sum_{m,n}Z_{m,n}\sin(m\vartheta-\nfp n\phi) , \nonumber
\end{eqnarray}
and the mode amplitudes $R_{m,n}$ and $Z_{m,n}$ are included in the parameter space. 

Fixed profiles of density and temperature are chosen, used both for the bootstrap current calculations and to set the pressure profile for the MHD equilibrium calculations.
For MHD equilibrium, one other profile must be specified, typically either the toroidal current $I_t(s)$ (or its radial derivative) or $\iota$. 
We find in practice that using the profile of $dI_t/ds$ in the parameter space is preferable as it yields slightly lower values of the objective than if $\iota(s)$ is used. 
However, varying $\iota(s)$ in the optimization is also effective.
The $dI_t/ds$ profile is represented as a cubic spline on fixed nodes, with the function values varied during the optimization. Since typical values of $dI_t/ds$ in SI units can be $\sim 10^6$, the optimizer performs better if the function values are scaled so they are $\sim 1$.

The optimization problem is solved using the SIMSOPT software \citep{simsopt}, with VMEC \citep{VMEC1983} as the equilibrium code. The nonlinear least-squares problem is solved using the default algorithm in scipy, ``trust region reflective''.
Gradients are provided by finite differences, using MPI for concurrent function evaluations.
The initial condition is a circular cross-section torus. For quasi-helical symmetry, torsion of the magnetic axis is added to the initial condition via $R_{1,1}$ and $Z_{1,1}$. 
The parameter space is expanded in a series of steps, with the surface shape and current profile refined at each step. In step $j=1, 2, \ldots$, the amplitudes $R_{m,n}$ and $Z_{m,n}$ with $m \le j$ and $|n| \le j$ are varied, and $2j+3$ spline nodes are used in the current profile. 
For quasi-axisymmetry, the approach is slightly different, and starts from the QA vacuum configuration presented in Ref.~\onlinecite{LandremanPaul2022} to ensure that $\iota$ is sufficient to support the pressure and that the local minimum is not a tokamak.
As a crude method to address the presence of local minima in the objective, for each case shown in the following sections, 12-24 optimizations are run with different finite difference step sizes, and the result with lowest objective function is shown.
The software, scripts, input files, and output files associated with the optimizations are available at Ref.~\onlinecite{zenodo}. 

\changed{
The optimization problem described here is limited to a single density profile and single temperature profile. These profiles would be chosen to be the target operating point of an experiment. However in a real experiment it would be important to consider how the configuration changes for other profiles, particularly for the low temperature at the start of the discharge. Ideally, one would like to optimize for confinement and $\iota$ across a range of profiles. However it is not clear how the procedure of this section could be generalized to account for multiple sets of possible profiles, since the variation of a configuration with profiles requires a choice of coils and free-boundary calculations, which are not considered here. For experimental scenarios in which the bootstrap current is different from its design value, such as the ramp-up phase at the start of a discharge,  associated differences in $\iota$ could perhaps be compensated to some extent by other sources of current (driven by radio-frequency waves or beams) or control coils. Understanding how to optimize a stellarator design across a range of plasma profiles is an important question for future research.
}


\section{Optimization results}
\label{sec:results}


\subsection{Quasi-helical symmetry at moderate $\beta$}
\label{sec:QH_medium_beta}

As a first demonstration of the new optimization method,
we consider quasi-helical symmetry in a reactor-scale configuration with four field periods and an aspect ratio $A_*=6.5$. We adopt the density and temperature profiles (\ref{eq:profiles}) with $n_{e,0}=2.2\times 10^{20}$ m$^{-3}$ and $T_{e,0}=T_{i,0}=10$ keV. 

\begin{figure}
  \centering
  \includegraphics[width=\columnwidth]{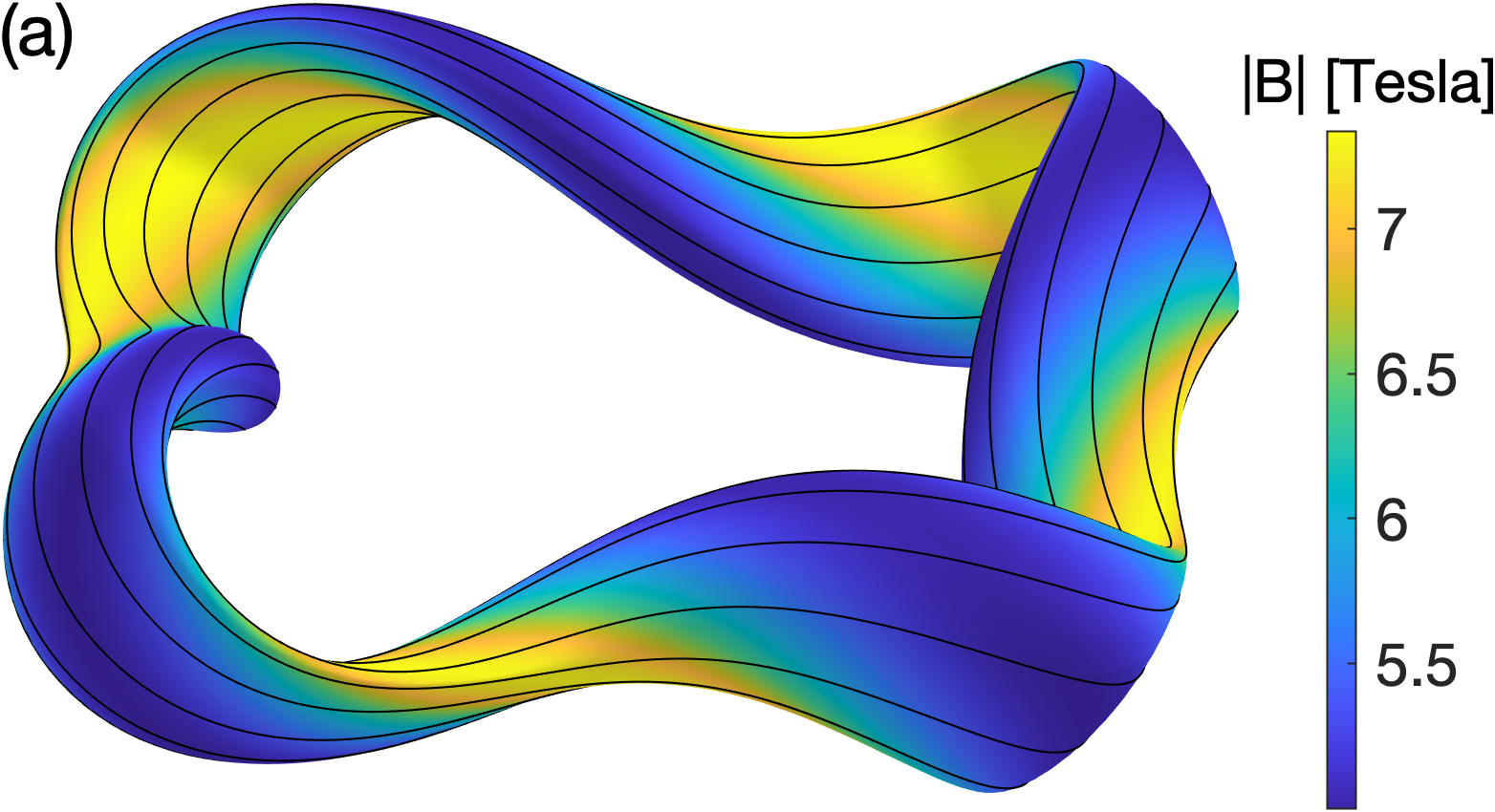}
  \hspace{0.1in}
  \includegraphics[width=2in]{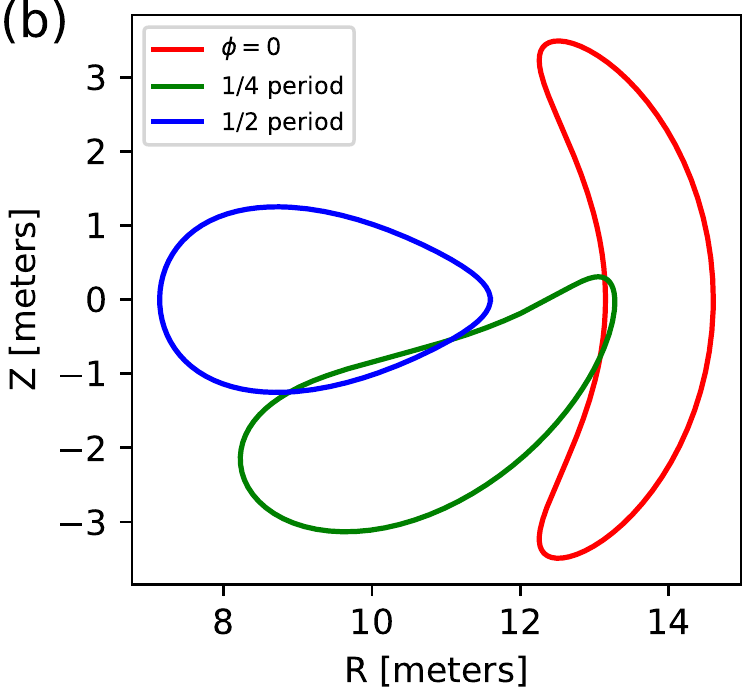}
  \caption{\label{fig:QH_medium_beta} 
Optimized quasi-helically symmetric configuration from section \ref{sec:QH_medium_beta}, with volume-averaged $\beta=2.5\%$.
\changed{(a) Three-dimensional view. (b) Cross sections at three toroidal angles.}
}
\end{figure}

\begin{figure}
  \centering
\includegraphics[width=\columnwidth]{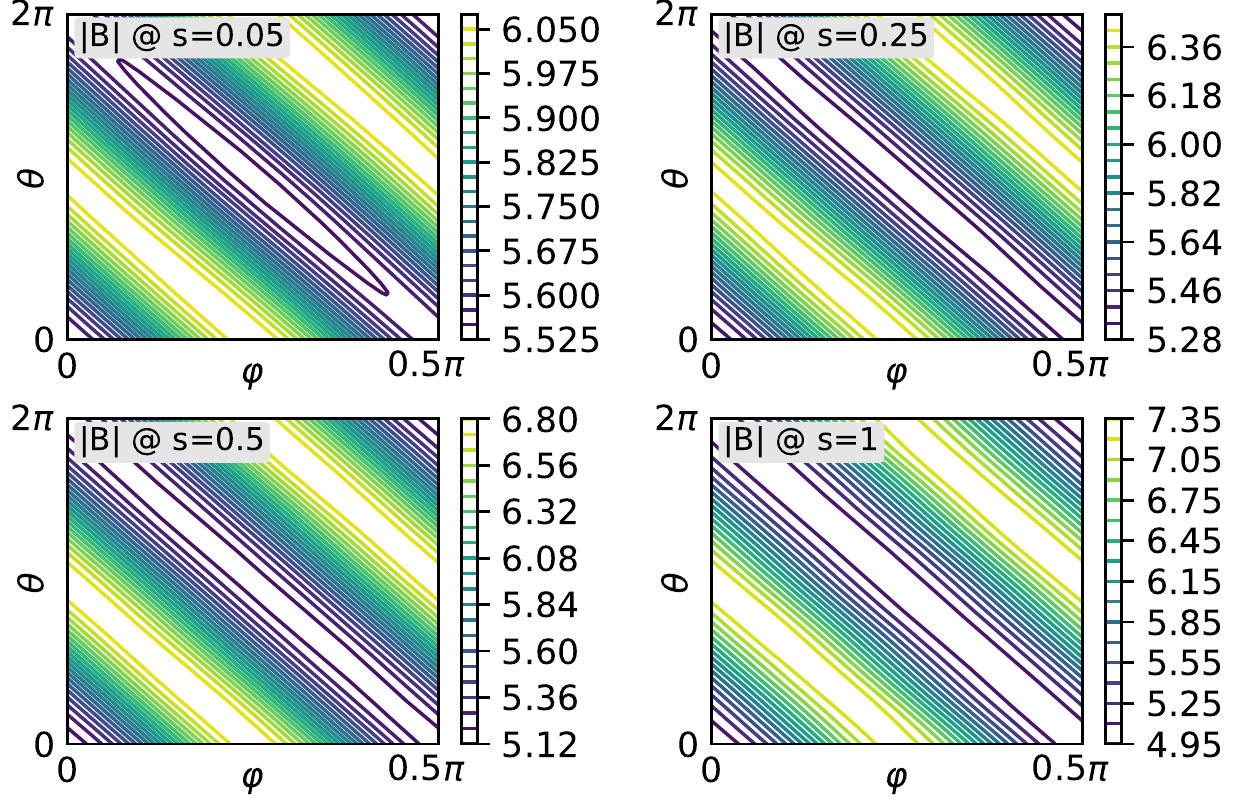}
  \caption{\label{fig:symmetry_medium_beta} 
Demonstration of good quasi-helical symmetry throughout the plasma volume for the configuration from section \ref{sec:QH_medium_beta}, with volume-averaged $\beta=2.5\%$.
}
\end{figure}

The configuration resulting from the optimization has a volume-averaged $\beta$ of 2.5\% and a plasma current of 1.2 MA. The geometry of the configuration is shown in figure \ref{fig:QH_medium_beta}. The geometry is qualitatively similar to previous quasi-helically symmetric stellarators \citep{HSX, KuQH, Wistell, LandremanPaul2022}. Contours of $B$ in Boozer coordinates are shown for several flux surfaces in figure \ref{fig:symmetry_medium_beta}, showing excellent quasisymmetry throughout the volume. Indeed, the neoclassical transport coefficient $\epsilon_{eff}^{3/2}$ is $<3\times 10^{-5}$, roughly 30 times smaller than in W7-X. For comparison, a recent estimate\cite{Alonso} suggests $\epsilon_{eff}^{3/2} \lesssim 10^{-3}$ is sufficiently small for a reactor, and for smaller values, turbulent transport will dominate.

\begin{figure}
  \centering
\includegraphics[width=\columnwidth]{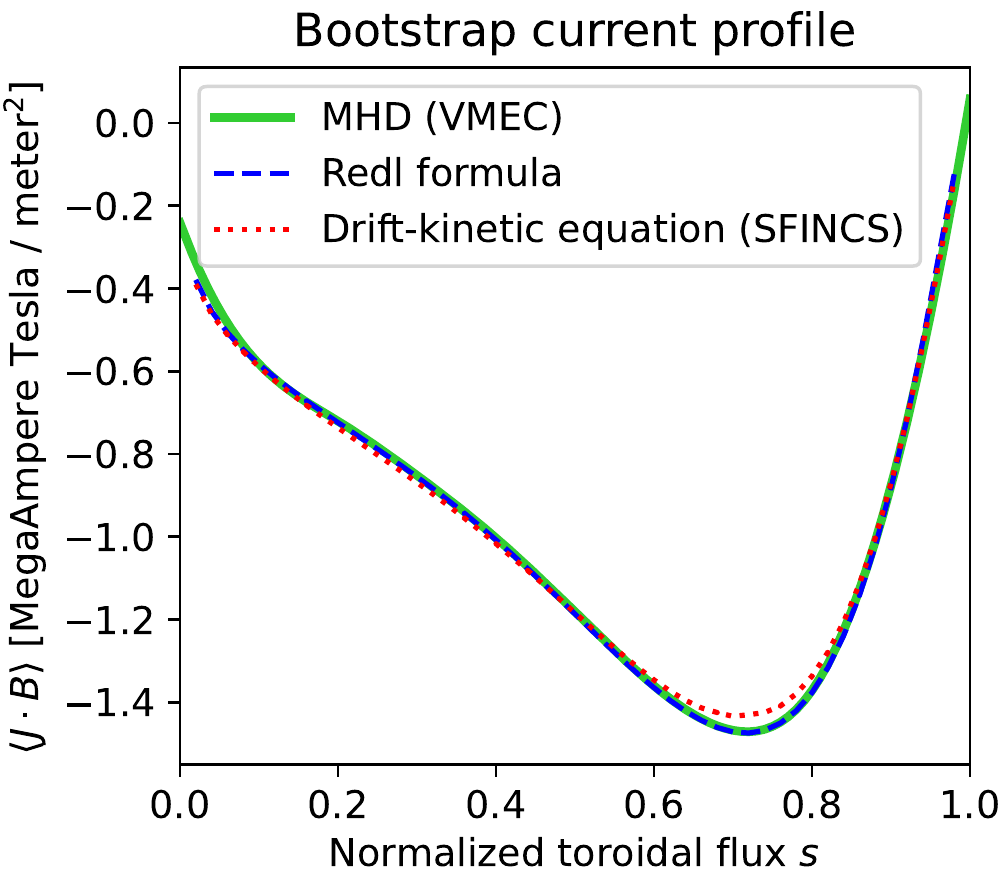}
  \caption{\label{fig:jdotB_QH_beta2p5} 
Self-consistency of the current profile for the quasi-helically symmetric configuration from section \ref{sec:QH_medium_beta}, with volume-averaged $\beta=2.5\%$.
}
\end{figure}

Figure \ref{fig:jdotB_QH_beta2p5} shows the profile of bootstrap current in the optimized configuration, computed three ways. The solid green curve shows the profile associated with the MHD equilibrium, taken from the VMEC output file.
The dashed blue curve shows the result of the Redl formula (as modified in section \ref{sec:redl}). These two curves overlap closely, as expected due to the minimization of $f_{boot}$. Finally, the dotted red curve shows a SFINCS calculation for the equilibrium, fully accounting for three-dimensional geometry and imperfections in the quasisymmetry, and considering the ions to be an equal mixture of deuterium and tritium. This curve also agrees well with the MHD equilibrium, showing that self-consistency of the current profile has been achieved.

\begin{figure}
  \centering
  \includegraphics[width=2.5in]{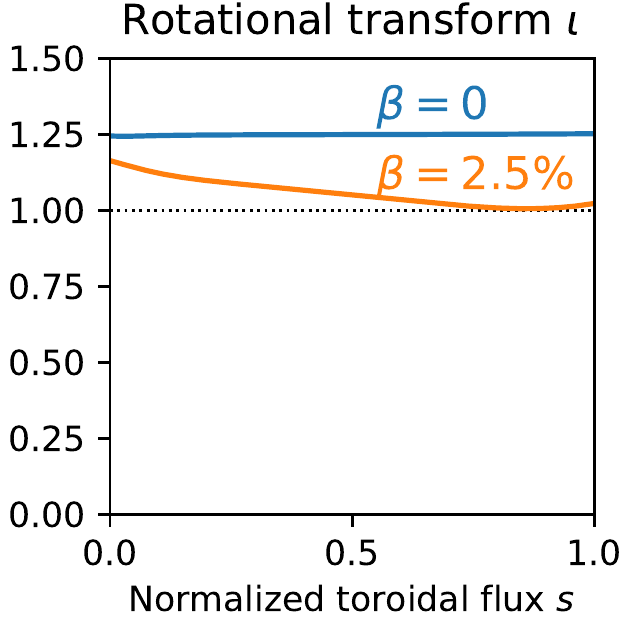}
  \caption{\label{fig:iota1} 
Profile of rotational transform $\iota$ for the quasi-helically symmetric configuration from section \ref{sec:QH_medium_beta}, with volume-averaged $\beta=2.5\%$. Compared to a similar optimization at $\beta=0$, it can be seen that the bootstrap current reduces $|\iota|$.
}
\end{figure}

A problem arises when this method is applied to larger values of $\beta$, which can be understood from figure \ref{fig:iota1}. Here, the profile of $\iota$ is shown both for the $\beta=2.5\%$ configuration and for a configuration optimized at the same aspect ratio at $\beta=0$. The bootstrap current in quasi-helical stellarators acts to reduce $|\iota|$, and for $\beta=2.5\%$, the $\iota$ profile is close to crossing $\iota=1$. This value is the worst possible rational value, in that tiny field errors are likely to introduce large magnetic islands. For larger values of $\beta$, a method of controlling $\iota$ must be introduced to avoid crossing $\iota=1$.

\subsection{Quasi-helical symmetry at high $\beta$}
\label{sec:QH_high_beta}

To optimize for quasi-helical symmetry at higher $\beta$, the following term is added to the objective function:
\begin{equation}
    f_\iota = \int_0^1 ds \left[ \min(|\iota|-1.03, \; 0) \right]^2.
    \label{eq:iota_barrier}
\end{equation}
This term acts as a barrier to prevent $\iota$ from dropping below 1.03, providing some margin away from the resonance at $\iota=1$.

\begin{figure}
  \centering
  \includegraphics[width=\columnwidth]{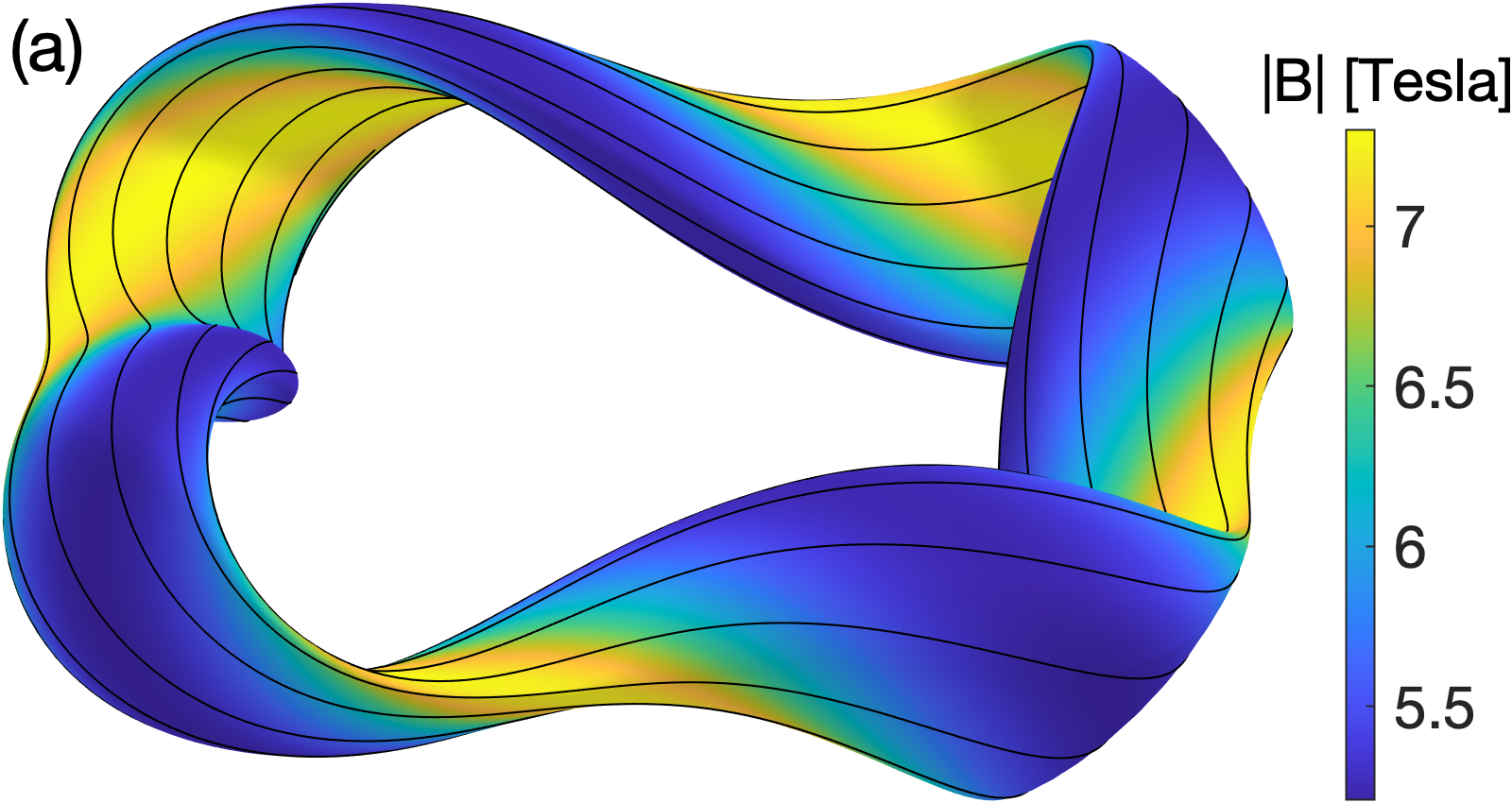}
  \hspace{0.1in}
  \includegraphics[width=2in]{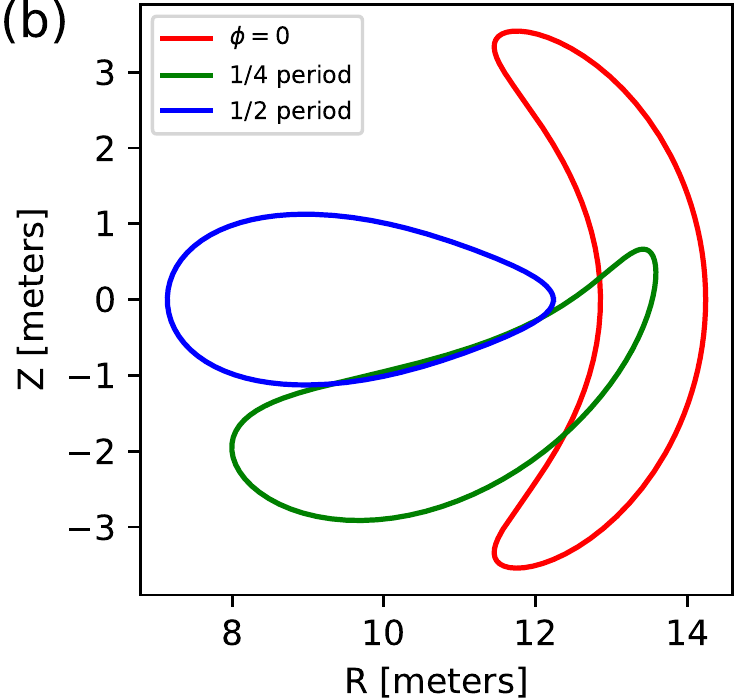}
  \caption{\label{fig:QH_high_beta} 
Optimized quasi-helically symmetric configuration from section \ref{sec:QH_high_beta}, with volume-averaged $\beta=5\%$, avoiding $\iota=1$.
\changed{(a) Three-dimensional view. (b) Cross sections at three toroidal angles.}
}
\end{figure}

\begin{figure}
  \centering
\includegraphics[width=\columnwidth]{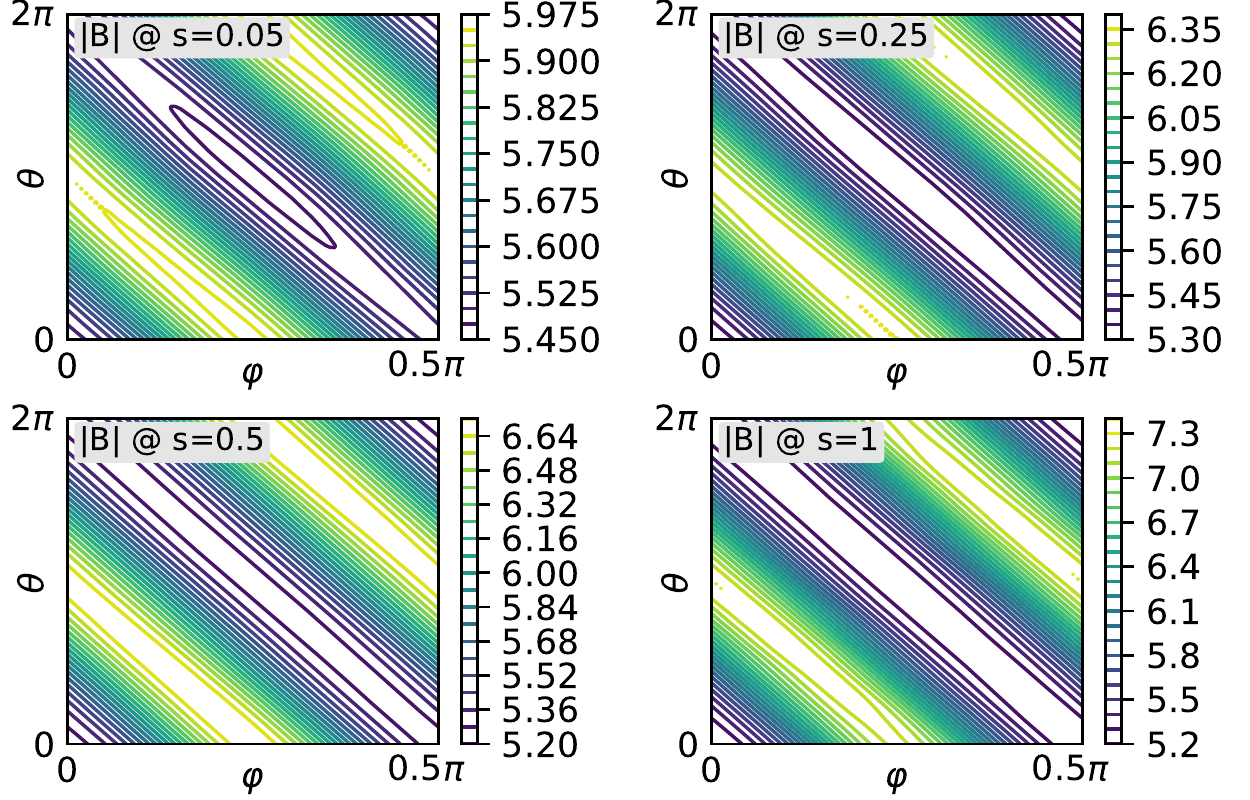}
  \caption{\label{fig:symmetry_high_beta} 
Demonstration of good quasi-helical symmetry for the configuration from section \ref{sec:QH_high_beta}, with volume-averaged $\beta=5\%$, before fixed-point iterations.
}
\end{figure}

An example of optimization using this barrier term at $\beta=5\%$ is shown in figure \ref{fig:QH_high_beta}.
For this optimization we again consider an aspect ratio $A_*=6.5$ and the profiles (\ref{eq:profiles}), now with $n_{e,0}=3\times 10^{20}$ m$^{-3}$ and $T_{e,0}=T_{i,0}=15$ keV. Excellent quasisymmetry throughout the volume is again achieved, shown by the contours of $B$ in Boozer coordinates in figure \ref{fig:symmetry_high_beta}.

\begin{figure}
  \centering
  \includegraphics[width=2.5in]{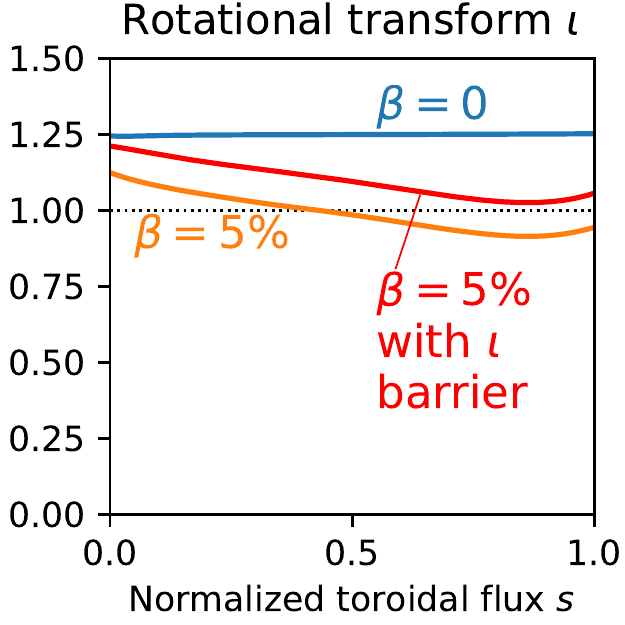}
  \caption{\label{fig:iota2} 
Profile of rotational transform $\iota$ for the quasi-helically symmetric configuration from section \ref{sec:QH_high_beta}, with volume-averaged $\beta=5\%$. A similar optimization without $f_\iota$ in (\ref{eq:iota_barrier}) and a vacuum configuration with otherwise identical parameters are also shown.
}
\end{figure}

Figure \ref{fig:iota2} shows that the rotational transform profile avoids the $\iota=1$ resonance. For comparison, a similar optimization with the same density and temperature profiles but without the barrier term (\ref{eq:iota_barrier}) is also shown. In this case the $\iota=1$ resonance is crossed in the middle of the volume. Interestingly, the barrier has the effect of shifting the $\iota(s)$ profile by a constant, without affecting the magnetic shear.

\begin{figure}
  \centering
\includegraphics[width=\columnwidth]{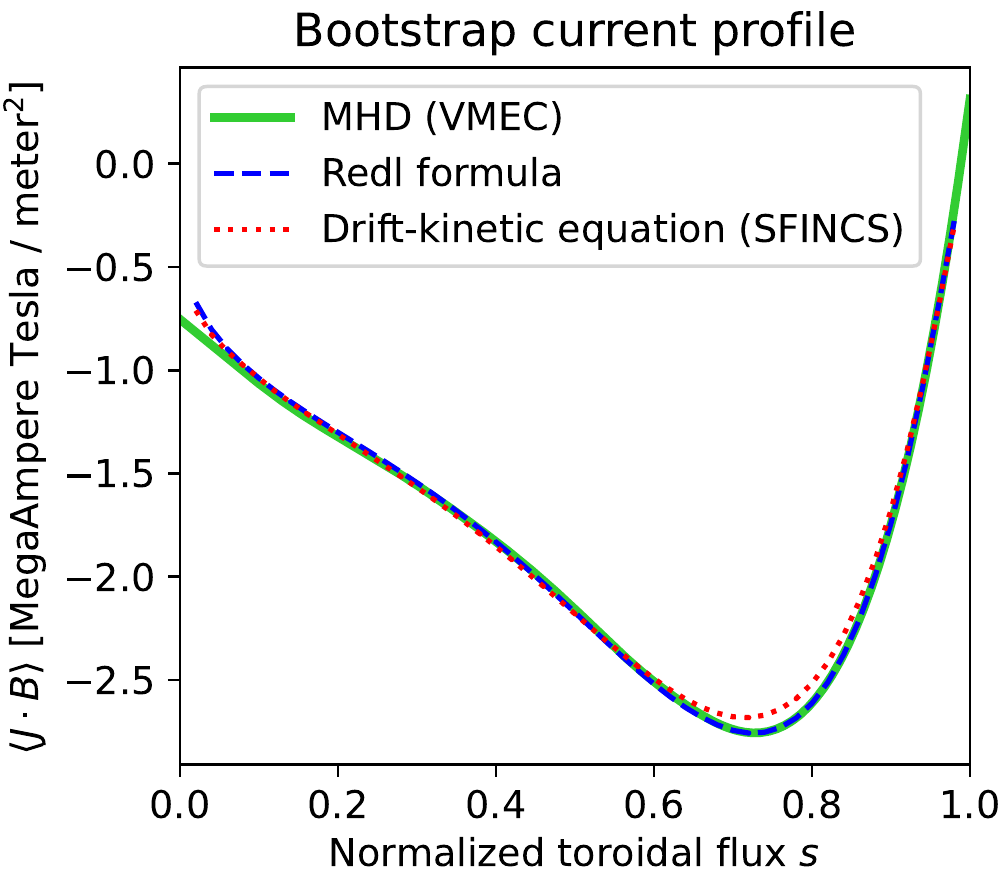}
  \caption{\label{fig:jdotB_QH_beta5} 
Self-consistency of the current profile for the quasi-helically symmetric configuration from section \ref{sec:QH_high_beta}, with volume-averaged $\beta=5\%$, before fixed-point iterations.
}
\end{figure}

Figure \ref{fig:jdotB_QH_beta5} shows the current profile for the final configuration, computed in the same three ways as in figure \ref{fig:jdotB_QH_beta2p5}. Again it can be seen that the current profile is self-consistent, with only a tiny difference between $\left<\vect{j}\cdot\vect{B}\right>$ in the MHD equilibrium compared to the SFINCS calculation.

If there is any concern about the small remaining differences between the MHD and drift-kinetic current profiles, the difference can be made arbitrarily small in a post-processing step. Following the optimization, a small number of fixed-point iterations between MHD equilibrium and the drift-kinetic equation can be applied, as described in the introduction. The boundary shape is fixed during these iterations. While there is no control on other terms in the objective function such as quasisymmetry or $f_\iota$ during the fixed-point iterations, the change to the current profile is extremely small, since the current profile was very nearly self-consistent before the fixed-point iterations. Therefore, degradation of the other objectives is negligible.

\begin{figure}
  \centering
\includegraphics[width=\columnwidth]{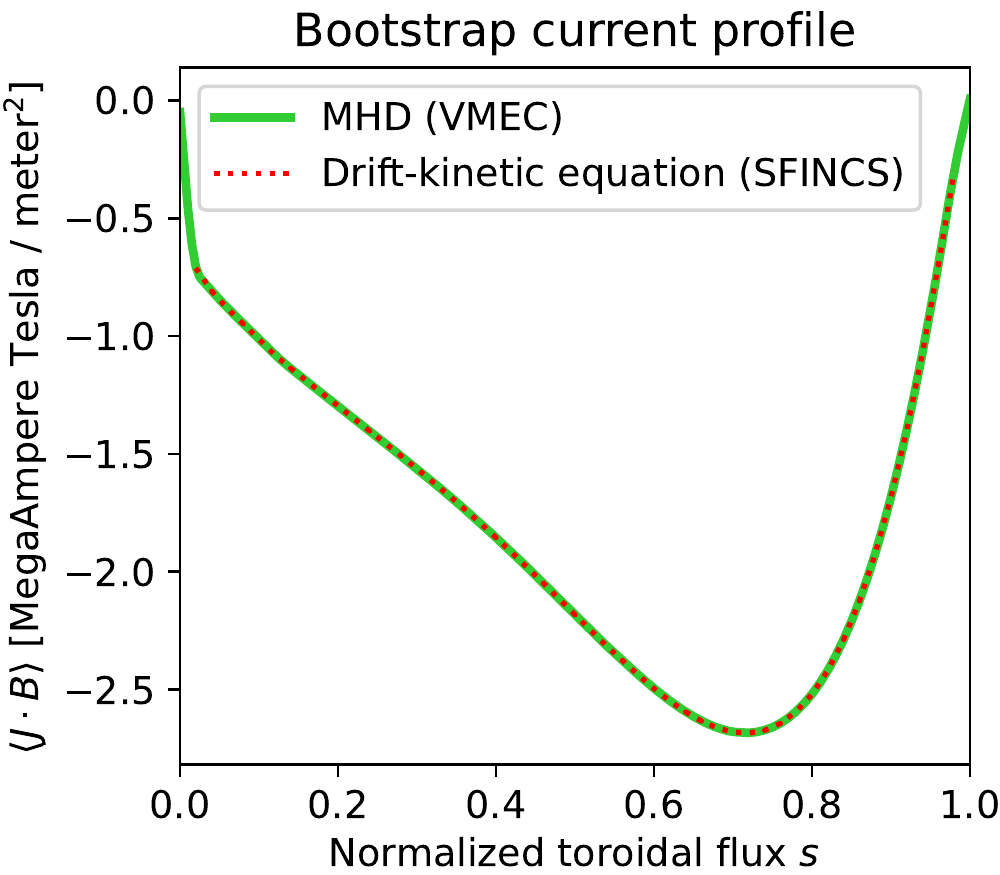}
  \caption{\label{fig:jdotB_after_fixed_point} 
Self-consistency of the current profile for the quasi-helically symmetric configuration from section \ref{sec:QH_high_beta}, with volume-averaged $\beta=5\%$, after fixed-point iterations.
}
\end{figure}

\begin{figure}
  \centering
\includegraphics[width=\columnwidth]{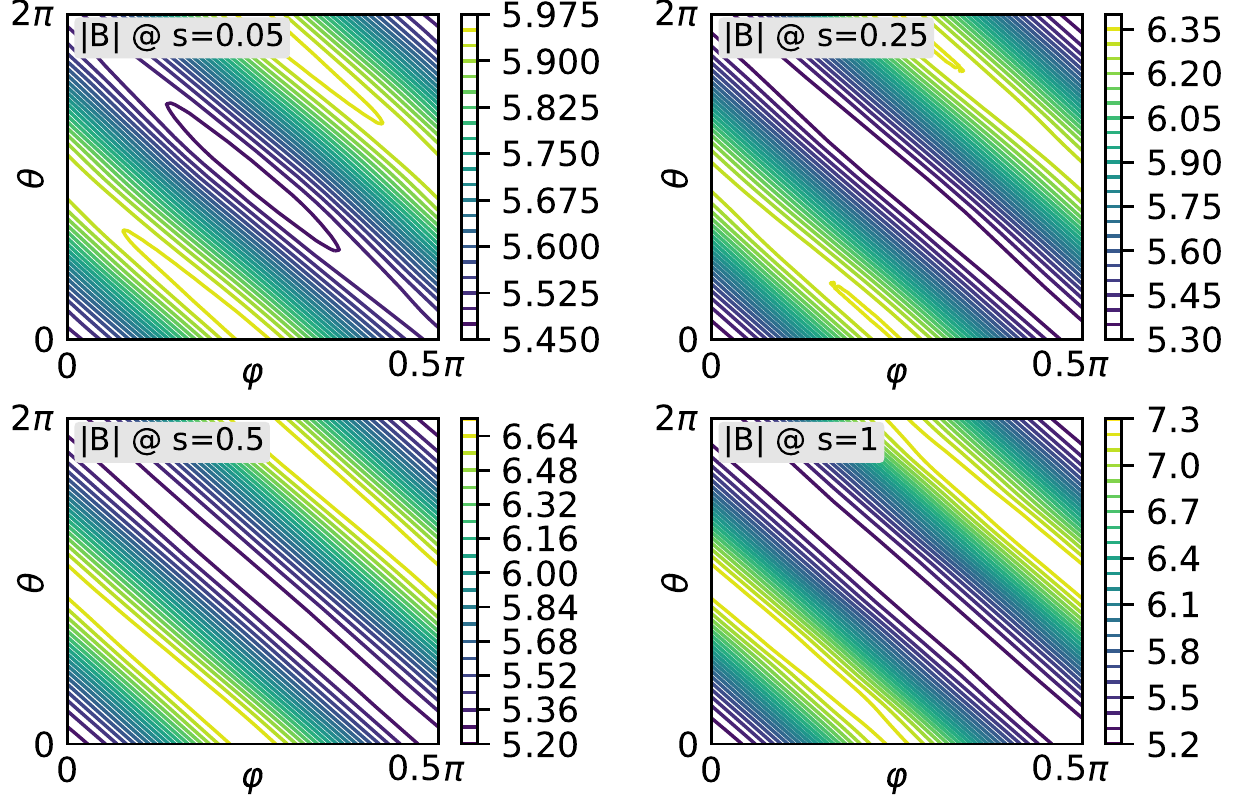}
  \caption{\label{fig:symmetry_high_beta_after_fixed_point} 
Demonstration of good quasi-helical symmetry for the configuration from section \ref{sec:QH_high_beta}, with volume-averaged $\beta=5\%$, after fixed-point iterations.
}
\end{figure}

Fixed-point iterations using SFINCS are applied to the configuration of figure \ref{fig:QH_high_beta}, and the final current profile is shown in figure \ref{fig:jdotB_after_fixed_point}. It can be seen that there is now essentially exact agreement between MHD equilibrium and the drift-kinetic equation.
Figure \ref{fig:symmetry_high_beta_after_fixed_point} shows $B$ as a function of Boozer angles after the fixed-point iterations. There is no significant degradation in quasisymmetry compared to figure \ref{fig:symmetry_high_beta}, before the fixed-point iterations. The final configuration has $\epsilon_{eff}^{3/2} < 6\times 10^{-5}$, so neoclassical transport will be negligible.

\subsection{Quasi-axisymmetry}
\label{sec:QA}
For QA configurations, we start the optimization from the ``new QA'' geometry in Ref.~\onlinecite{LandremanPaul2022}. This is an $\nfp = 2$ vacuum configuration with an average $\iota$ of $0.42$, and aspect ratio $A=6.0$. Like the other configurations in this paper, we have $a=1.70$ m and $\bar{B}=5.86$ T. We use profiles (\ref{eq:profiles}) with $n_{e,0} = 2.38 \times 10^{20}$ m$^{-3}$ and $T_{e,0} = T_{i,0} = 9.45$ keV. The target average $\iota$ is achieved by including a term (\ref{eq:QA_mean_iota_target}) in the optimization.
To prevent $\iota$ from crossing $0.5$, a barrier term as in (\ref{eq:iota_barrier}) is employed, with the barrier at $\iota = 0.485$ and a maximum employed instead of a minimum to penalize crossing \emph{above} this $\iota$. No stepping in the number of Fourier-modes is done from this initial condition, as this results in the optimizer finding an axisymmetric tokamak as the local minimum.

\begin{figure}
  \centering
    \includegraphics[width=\columnwidth]{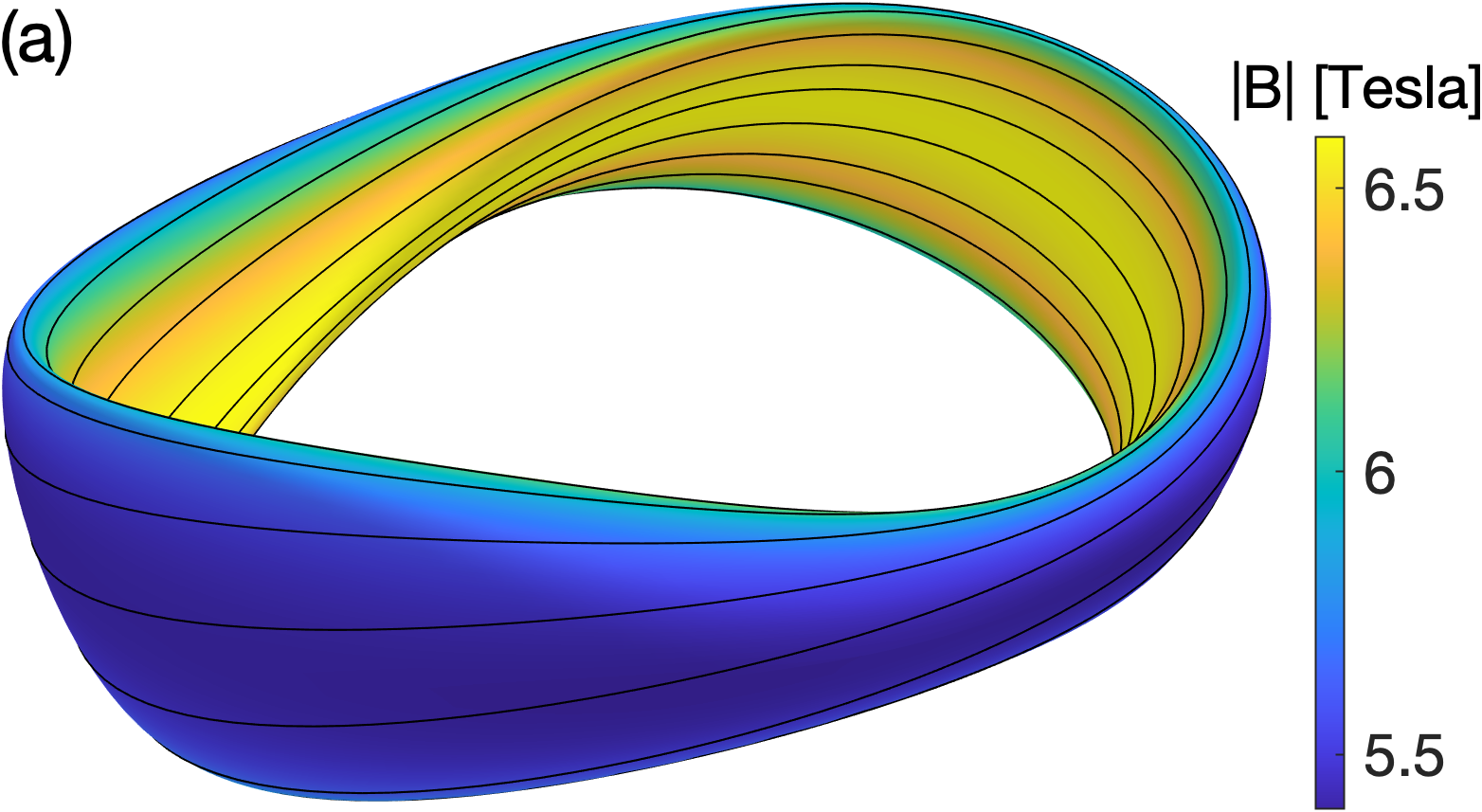}
  \hspace{0.1in}
  \includegraphics[width=2in]{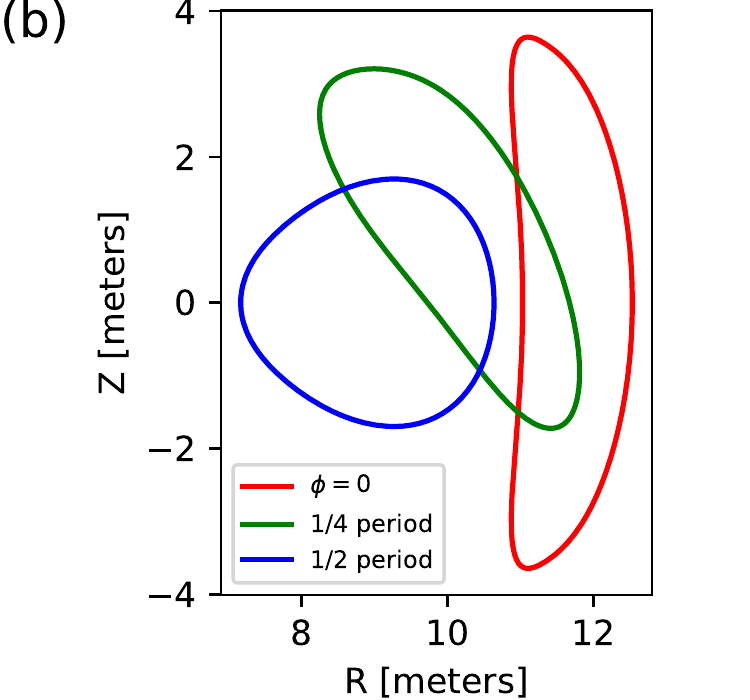}
  \caption{\label{fig:QA_beta2p5} 
Optimized quasi-axisymmetric configuration from section \ref{sec:QA}, with volume-averaged $\beta=2.5\%$.
\changed{(a) Three-dimensional view. (b) Cross sections at three toroidal angles.}
}
\end{figure}

\begin{figure}
  \centering
\includegraphics[width=\columnwidth]{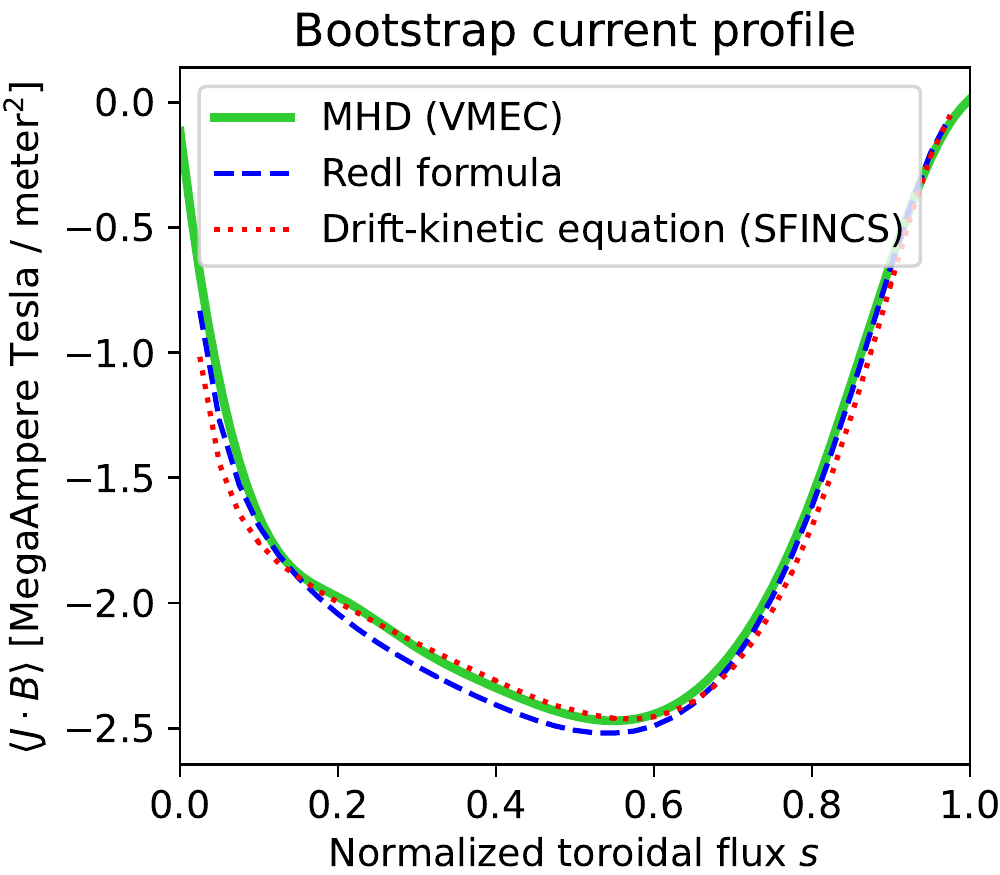}
  \caption{\label{fig:jdotB_QA_beta2p5} 
Self-consistency of the current profile for the quasi-axisymmetric configuration from section \ref{sec:QA}, with volume-averaged $\beta=2.5\%$.
}
\end{figure}

\begin{figure}
  \centering
\includegraphics[width=\columnwidth]{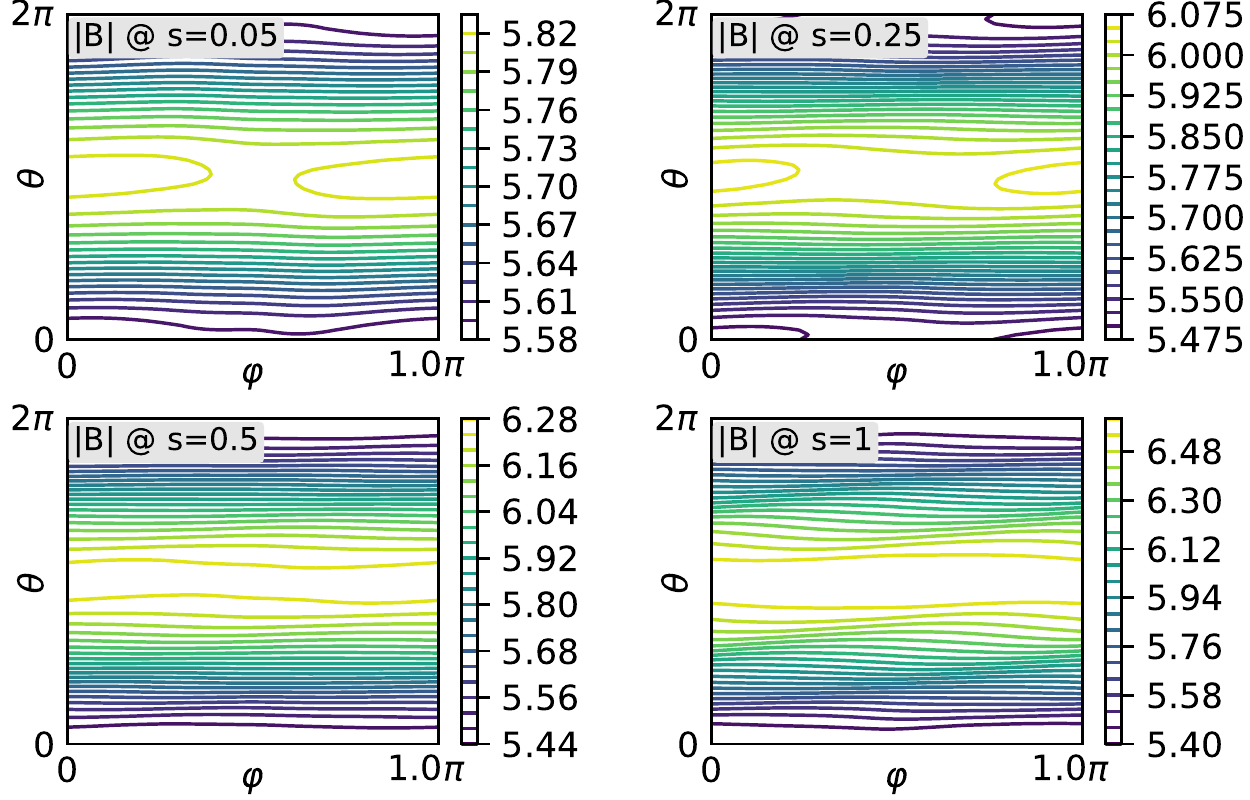}
  \caption{\label{fig:symmetry_QA_beta2p5} 
Demonstration of good quasi-axisymmetry for the configuration from section \ref{sec:QA}, with volume-averaged $\beta=2.5\%$.
}
\end{figure}

\begin{figure}
  \centering
  \includegraphics[width=2.5in]{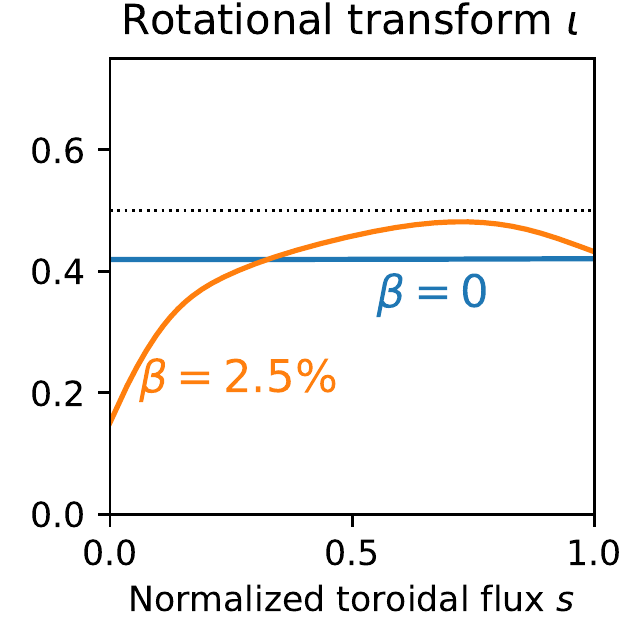}
  \caption{\label{fig:iota_QA} 
Profile of rotational transform $\iota$ for the quasi-axisymmetric configuration from section \ref{sec:QA}.
}
\end{figure}

The resulting configuration has a volume averaged $\beta$ of $2.5\%$, and a $2.72$ MA plasma current. 
\changed{
This value is $2.3\times$ larger than the plasma current for the QH configuration at the same $\beta$ in section \ref{sec:QH_medium_beta}, consistent with all neoclassical effects being larger in QA compared to QH symmetry.}
The neoclassical transport coefficient $\epsilon_{eff}^{3/2}$ is $<2\times 10^{-5}$, which again is more than adequate.
The surface shape, bootstrap current profile, the quality of the quasi-axisymmetry, and the $\iota$ profile are shown in figures \ref{fig:QA_beta2p5} - \ref{fig:iota_QA}.
The figures show that the boundary has modest shaping, the Redl formula is accurate, reasonable quasi-axisymmetry is achieved, and the $\iota$ profile avoids $1/2$.

The $\iota$ profile does however cross several other low-order rationals, including $1/4$, $2/5$, and $1/3$, but these crossings are confined to the core region where the magnetic shear is relatively large, reducing the width of islands even if resonant fields are present. It may be difficult to entirely avoid crossing low-order rational surfaces in QA stellarators at finite $\beta$ with self-consistent bootstrap current, so having these crossings in the core might be the best achievable outcome. The high shear across these rationals also limits the width of the region with low $\iota$, which is beneficial for fast particle confinement as high $\iota$ reduces the orbit width.

The downside of higher $\iota$ is that it increases elongation.
For vacuum QA configurations, this effect appears to limit $\iota$ to values of $\iota \lesssim 0.9$, based on previous studies exploring the landscape of quasisymmetric stellarators \cite{Rodriguez2022arXiv}.
In practice, the space of vacuum QA configurations may be yet more limited, as the plasma boundary tends to become unreasonably elongated already for mean $\iota$ of about $0.6$.

Adding finite $\beta$ to these configurations, in the same manner as was done here with using the vacuum configuration as an initial guess in the optimization, does not make the plasma boundaries less elongated in our optimizations.

In general, trying to increase $\beta$ much beyond 2.5\% for quasi-axisymmetric configurations, trying various choices of $\bar{\iota}_*$, initial conditions, and optimization sequences,
we have so far encountered three different possible modes of failure. (1) The optimal configuration can have an unreasonably elongated boundary\changed{, e.g. the ratio of boundary surface height to width can exceed 10 in the bean-shaped cross-section. This is probably unacceptable since the alpha particle Larmor radius can become a sizeable fraction of the shorter dimension, and the very large pressure gradient in real space could drive instabilities.}
(2) The optimal configuration can become close to a tokamak, with most of $\iota$ being created by the bootstrap current. (3) The configuration may not be numerically converged: high-resolution VMEC calculations may show interior flux surface shapes that appear unphysical, being far from elliptical in the core.
In our optimizations, failure state (1) is the most common. Failure state (2) is less common for larger values of $\bar{\iota}^*$.

It remains possible that for some other choices of parameters and objective function,
configurations with self-consistent bootstrap current and a similar degree of quasi-axisymmetry may exist at higher $\beta$ and/or at higher $\iota$.
In the future, other functions in the objective could be explored for avoiding axisymmetric solutions and solutions with large elongation.


\section{Comparison to long-mean-free-path calculations}
\label{sec:bootsj}

Previously, a different approach for computing the bootstrap current in stellarators has been to use a formula derived by expanding in low collisionality, i.e. long mean-free-path compared to system size \citep{Ware,Cfqsbootstrap}. Variations of this formula have been derived by several authors \citep{ShaingCallen,Shaing89,HelanderBootstrap17}. Due to the low-collisionality expansion, these formulae allow the bootstrap current to be computed faster than if the finite-collisionality 3D drift-kinetic equation is solved (as in the codes SFINCS and DKES \citep{DKES}). However, for quasisymmetric stellarators, the Redl formula is superior to these low-collisionality formulae for several reasons: the Redl formula is faster to evaluate, it is more accurate, and the low-collisionality formulae are plagued by noise.

\begin{figure}
  \centering
\includegraphics[width=\columnwidth]{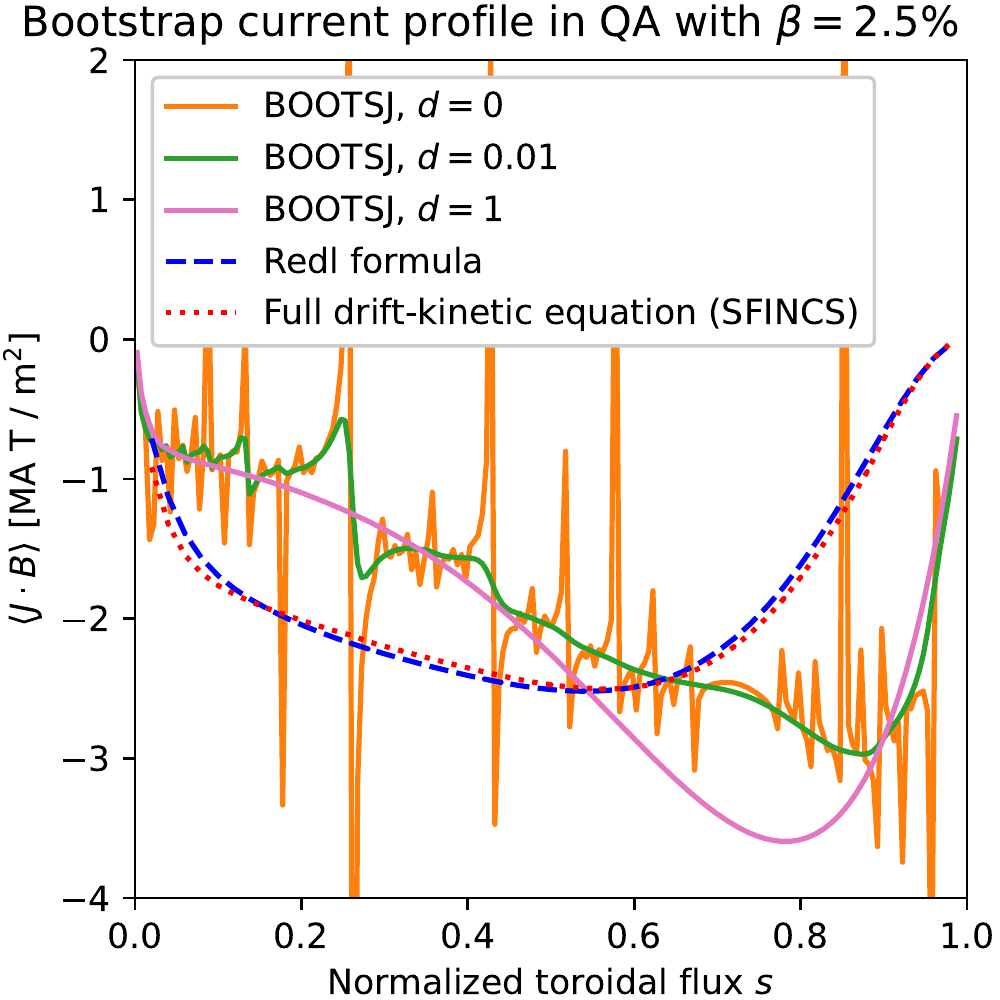}
  \caption{\label{fig:bootsj} 
The long-mean-free-path asymptotic formula for the bootstrap current, computed here with the code BOOTSJ, is both noisy and inaccurate. In contrast, the Redl formula is accurate and does not require an ad-hoc smoothing parameter $d$.
}
\end{figure}

These issues are shown in figure \ref{fig:bootsj}. Here, three methods for computing the bootstrap current are compared for the quasi-axisymmetric configuration of section \ref{sec:QA} with $\beta=2.5\%$. The red dotted curve shows a calculation of the full 3D drift-kinetic equation with SFINCS, which is the most complete physics model. The Redl formula is shown in the blue dashed curve, displaying close agreement with SFINCS. The orange curve shows the low-collisionality bootstrap formula computed with the code BOOTSJ \cite{Shaing1986,Shaing89,ShaingCrume}. This curve is extremely noisy, looking nothing like the Redl or SFINCS results. BOOTSJ includes an ad-hoc smoothing parameter named ``damp\_bs'', denoted here by $d$. This parameter is defined as follows: resonances $1/x$ in the bootstrap formula are replaced with $x/(x^2 + m^2 d^2)$, where $x=m-n/\iota$, and $m$ and $n$ are poloidal and toroidal mode numbers in the solution. 
The green and pink curves in figure \ref{fig:bootsj} show BOOTSJ results for $d=0.01$ and $d=1$. It can be seen that even though the spikes of the orange curve can be smoothed over with a sufficiently large $d$, the current profile still does not match the SFINCS or Redl results. The reason \changed{may be in part} that the bootstrap current approaches the low-collisionality asymptote extremely slowly, as shown for example in figure 3b of Ref.~\onlinecite{KernbichlerNeo2}. In contrast the Redl formula is valid for arbitrary collisionality.
\changed{Another contributing factor to the difference may be finite aspect ratio effects not included in BOOTSJ.}

In summary, for computing the bootstrap current in quasisymmetric stellarators, the Redl formula has many advantages over low-collisionality asymptotic formula.

\section{Confinement}
\label{sec:confinement}

\begin{figure*}
  \centering
\includegraphics[width=7in]{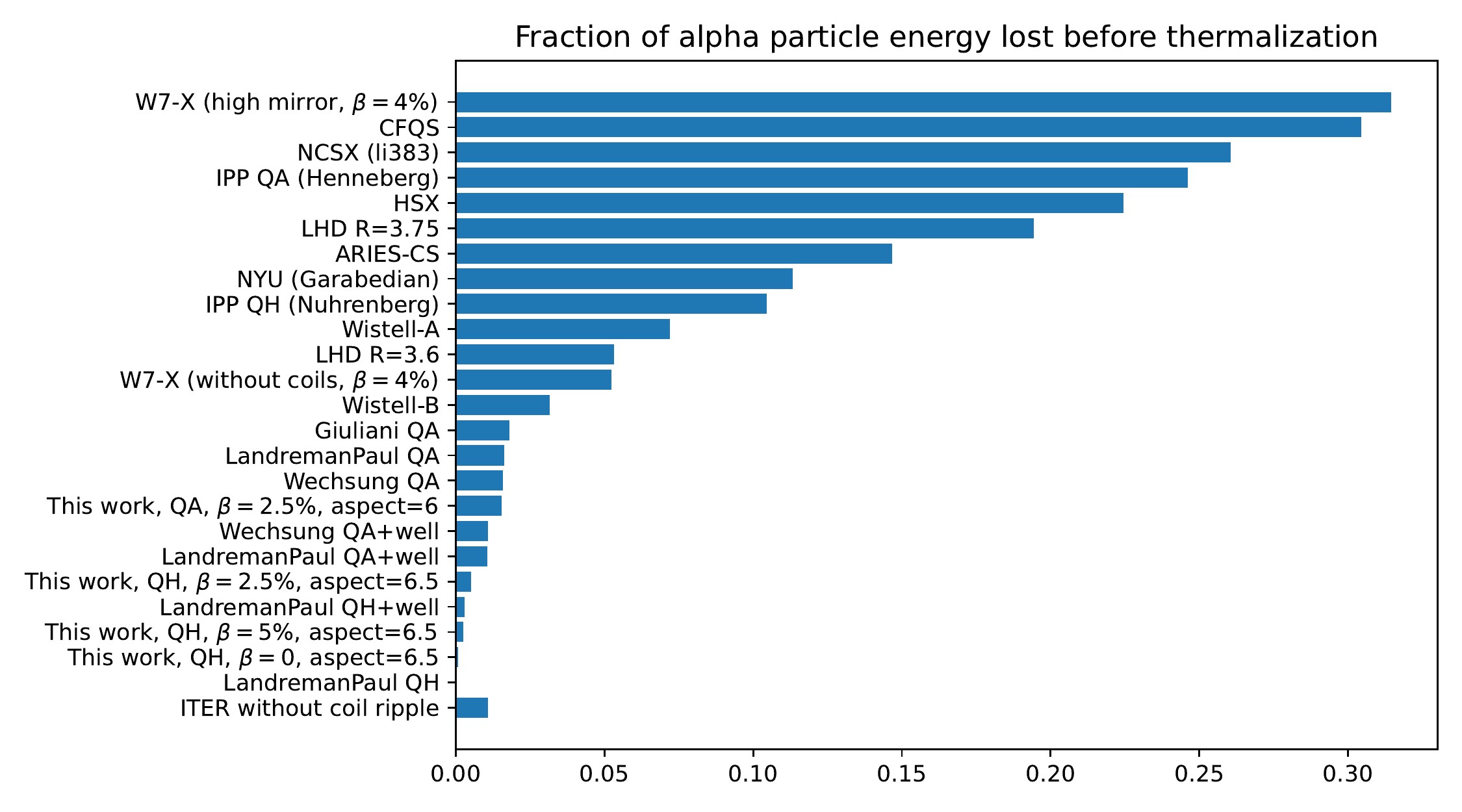}
  \caption{\label{fig:alphas} 
Significant progress has been made in the confinement of energetic particles in stellarators.
Loss of alpha particle energy is shown for a variety of magnetic configurations, all scaled to the same minor radius and average field strength.
}
\end{figure*}

Confinement of fusion-produced alpha particles has long been a serious challenge for the stellarator concept.
\changed{Several QA and QH optimizations have been recently demonstrated to achieve extremely good confinement of fast ions \cite{LandremanPaul2022,WechsungPNAS,WechsungAPriori,giulianiSurfaces} in vacuum fields, but a natural question is whether similarly good confinement is possible at finite beta.
We can now answer this question in the affirmative:}
the stellarator configurations obtained in the preceding sections have excellent alpha confinement, as a result of the high degree of quasisymmetry. Both the problem and the recent progress towards its resolution are shown in figure \ref{fig:alphas}.
This chart displays the fraction of alpha particle energy that is lost for a variety of stellarator configurations, using calculations similar to the recent study in Ref.~\onlinecite{Bader2021}.
Each configuration has been scaled to the minor radius and averaged field strength of ARIES-CS.
In each configuration, 20,000 alpha particles are initialized isotropically in velocity space at 3.5 MeV, with a spatial density proportional to the local fusion reaction rate. The latter is determined using the same profiles for the density and temperature of deuterium and tritium in each configuration, $n_D=n_T=(2\times 10^{20}\; \mathrm{m}^{-3})(1-s^5)$
and $T_D = T_T = (12 \, \mathrm{keV})(1-s)$.
The alpha guiding centers are followed with collisions until a particle either thermalizes with the bulk plasma or moves outside the $s=1$ surface, whereupon it is considered lost (a conservative assumption). The calculations are performed with the ANTS code \citep{Drevlak2014}. ANTS uses the relativistic guiding center equations on pages 42-43 of Ref.~\onlinecite{Sivukhin}.
Test particle collisions with the bulk plasma are computed using the same profiles as for the alpha birth distribution, along with an electron species with $n_e = n_D + n_T$ and $T_e = T_D$. These profiles are not generally consistent with the pressure profile used to compute each MHD equilibrium. But to fairly compare alpha confinement across configurations with varied $\beta$ it is more important that the birth and collision profiles be matched.

The set of magnetic configurations includes
W7-X \citep{W7X}, NCSX \citep{NCSX}, ARIES-CS \cite{ARIESCS}, CFQS \citep{CFQS}, HSX \cite{HSX}, and LHD \citep{LHD}. 
For LHD, since it is known that confinement is sensitive to shifts in the major radius, configurations with two values of major radius are included.
Also included are configurations from Refs.~\onlinecite{NuhrenbergZille,GarabedianNIST,Henneberg,Wistell,  Bader2021,LandremanPaul2022,WechsungPNAS,giulianiSurfaces}.

Figure \ref{fig:alphas} shows that for the majority of the stellarators designed before the year 2021, alpha energy losses of tens of percent are typical. These losses would likely cause significant damage to the plasma facing materials. For comparison, a perfectly axisymmetric ITER hybrid configuration is included at the bottom of the figure, with losses of only $\sim 1\%$. (Although this ITER configuration is perfectly symmetric, losses are still possible due to direct orbit loss and collisional diffusion.)
It is interesting that the LHD configuration with smaller major radius has lower losses than many stellarators that are nominally  optimized for reduced neoclassical transport.
However the losses are significantly lower, $\sim 2\%$ or smaller, for the configurations from section \ref{sec:results} of this paper,
along with the other configurations reported in the past year.
It is interesting that the QH configurations can have losses lower than the ITER configuration. This is due to the fact that the width of banana orbits in an axisymmetric or quasisymmetric configuration scales as $\propto 1/|\iota-N|$. Therefore the banana orbits are significantly thinner in the QH configurations, leading to reduced alpha loss. Note that the finite width of banana orbits is independent of the bounce-averaged drifts, so confinement measures based on bounce averages such as $\Gamma_c$ in Ref.~\onlinecite{Nemov} will not account for this difference between QA and QH configurations.

Comparing alpha particle confinement between stellarators and tokamaks is complicated, so the ITER bar in figure \ref{fig:alphas} must be interpreted with caution. On the one hand, higher field strengths may be possible in tokamaks, since greater structural support would be needed in a stellarator to prevent bending deformations of non-planar coils under electromagnetic loads. This difference favors better alpha confinement in tokamaks. On the other hand, stellarators can operate at higher collisionality, due to the absence of the Greenwald density limit, and the shorter slowing-down time would favor confinement in stellarators.


\section{Discussion and conclusions}
\label{sec:conclusions}

In this work we have demonstrated a new method for computing the bootstrap current in quasisymmetric stellarators, leveraging the isomorphism with tokamaks. The method enables the optimization of quasisymmetric stellarators such that the current profile in the MHD equilibrium is consistent with the drift-kinetic equation.
Since the bootstrap current is computed in this method using explicit analytic formulae,
the computational cost of the bootstrap calculations is negligible compared to the MHD equilibrium calculations. The new approach has the additional advantage that the current profile is smooth, in contrast to the noisy 
low-collisionality asymptotic formulae for the bootstrap current in general stellarators.

For equal minor radius, field strength, and profiles of density and temperature, the new configurations have alpha particle confinement comparable to or even better than a tokamak. Similarly excellent confinement of alpha particles is seen for other $\beta=0$ stellarators reported in recent months \citep{LandremanPaul2022,WechsungPNAS,giulianiSurfaces}.

Although quasisymmetry imperfections in the configurations here are
consistently smaller than in pre-2021 configurations,
the imperfections 
at finite $\beta$ are larger than in vacuum configurations at the same aspect ratio. In future work, it would be valuable to understand this phenomenon. One explanation could be the following. As discussed in Ref.~\onlinecite{GB2}, in an expansion in small distance from the magnetic axis compared to the scale length of the axis shape, quasisymmetry is likely to be possible to two orders but not three. Therefore, quasisymmetry is likely best when gradients in the minor radius direction are small, so the third-order terms are as small as possible. The bootstrap current profile introduces relatively short-scale variation in the minor radius direction, potentially increasing the size of the symmetry-breaking third-order terms. Supporting this theory, preliminary numerical experiments show that for fixed total toroidal current, quasisymmetry errors are reduced if an $s$-independent current profile is prescribed, in contrast to the self-consistent bootstrap current profiles used here. In the future, further investigations of the $\beta$-dependence of quasisymmetry errors would be worthwhile.

Significant additional work would be needed before the configurations in this paper could be considered complete conceptual designs.
First, MHD stability would need to be assessed, and possibly optimized. Second, flux surface quality would need to be checked using an MHD or extended-MHD code that does not assume the existence of flux surfaces. While both of the QH configurations avoid the low-order rationals at $\iota=1$ and $4/3$, they both cross $\iota=8/7 \approx 1.14$ and $12/11 \approx 1.09$, and the $\beta=5\%$ case crosses $\iota = 12/10=1.2$. Islands could exist at any of these resonances. Third, coil feasibility would need to be studied.

The most uncertain aspects of these configurations are the density and temperature profiles. In this work we have considered only prescribed profiles. In a real experiment these profiles would largely be determined by the sources of heat and particles together with turbulent transport, which is challenging to model. Due to this significant uncertainty, the sensitivity of the confinement to different density and temperature profiles should be studied.

\changed{
Another question for future work is the extent to which, within the constraint of QA or QH symmetry, the bootstrap current magnitude can be minimized.
Minimization of the current would help to preserve the other optimized properties over a wider range of density and temperature profiles, and reduce a potential source of MHD instability. While the quasisymmetry isomorphism implies that the bootstrap current in QA or QH symmetry cannot be eliminated, some freedom does exist in its magnitude. For given density and temperature profiles, the bootstrap current can be varied through $\iota$ and $f_t$, the latter of which can in principle be controlled through the ratio $B_{\max} / B_{\min}$ on the surfaces.
}


\begin{acknowledgments}
Magnetic configurations in figure \ref{fig:alphas} were provided by Andrew Giuliani,
Sophia Henneberg, Geoffrey McFadden, Shoichi Okamura, Neil Pomphrey, John Schmitt, Don Spong, Yasuhiro Suzuki, and Florian Wechsung.
Thanks also to Andreas Redl and Aaron Bader for providing clarifying input related to this research.
Support from the SIMSOPT team is gratefully acknowledged.
This work was supported by the
U.S. Department of Energy
under contracts DE-AC02-09CH11466 and DE-FG02-93ER54197.
This work was also supported by a grant from the Simons Foundation (560651, ML).
\end{acknowledgments}

\section*{Data Availability Statement}
The data that support the findings of this study are openly available in Zenodo at \url{https://doi.org/10.5281/zenodo.6520103},
Ref.~\onlinecite{zenodo}. 


\appendix

\section{The Redl formula}
\label{sec:redlAppendix}
In this section, to avoid confusion with the conventional tokamak notation for the poloidal flux, we denote the toroidal flux as $\psi_t$. Assuming zero induced electric field ($E_{\|} = 0$) and applying the substitutions in section \ref{sec:redl}, the Redl formula for a quasisymmetric stellarator is, from equation (2) in Ref.~\onlinecite{Redl},
\begin{equation}
\begin{aligned}
  \lang \vect{j} \cdot \vect{B} \rang = - \frac{\tilde{G}}{\iota - N}\left(\mcl_{31} \left[p_{\text{e}}\dndp + p_{\text{i}}\dnidp\right] \right.\\
  \left.
  + p_{\text{e}}(\mcl_{31} + \mcl_{32}) \dTedp + p_{\text{i}}(\mcl_{31} + \mcl_{34} \alpha) \dTidp\right).
  \end{aligned}
  \label{eq:redl}
\end{equation}
The coefficients are from equations (10-16) and (19-21) in Ref.~\onlinecite{Redl}:
\begin{widetext}
\begin{align}
  &\begin{aligned}\mcl_{31} = &\left(1 + \frac{0.15}{\zeff^{1.2} - 0.71} \right) \fteff{31} - \frac{0.22}{\zeff^{1.2} - 0.71}(\fteff{31})^2 \\ &+ \frac{0.01}{\zeff^{1.2} - 0.71}(\fteff{31})^3 + \frac{0.06}{\zeff^{1.2} - 0.71}(\fteff{31})^4,
               \end{aligned}
  \\
  &\mcl_{32} = F_{32,\text{ei}} + F_{32,\text{ee}}, \\
  &\mcl_{34} =  \mcl_{31}, \\
  &\begin{aligned}F_{32,\text{ee}} =&\frac{0.1 + 0.6\zeff}{\zeff (0.77 + 0.63[1 + (\zeff - 1)^{1.1}])}(\fteff{32,\text{ee}} - (\fteff{32,\text{ee}})^4) \\&+ \frac{0.7}{1 + 0.2\zeff} ((\fteff{32,\text{ee}})^2  -(\fteff{32,\text{ee}})^4 - 1.2[(\fteff{32,\text{ee}})^3- (\fteff{32,\text{ee}})^4]) \\&+ \frac{1.3}{1+0.5\zeff}(\fteff{32,\text{ee}})^4,
                      \end{aligned}
  \\
  &\begin{aligned}F_{32,\text{ei}} =&-\frac{0.4 + 1.93\zeff}{\zeff(0.8 + 0.6\zeff)} (\fteff{32,\text{ei}} - (\fteff{32,\text{ei}})^4) \\&+ \frac{5.5}{1.5 + 2\zeff} ((\fteff{32,\text{ei}})^2 - (\fteff{32,\text{ei}})^4 - 0.8[(\fteff{32,\text{ei}})^3 - (\fteff{32,\text{ei}})^4]) \\&- \frac{1.3}{1 + 0.5\zeff}(\fteff{32,\text{ei}})^4,
   \end{aligned}
\end{align}
\begin{align}
  &\frac{\ft}{\fteff{31}} = 1 + \frac{0.67(1 - 0.7\ft)\sqrt{\nue}}{0.56 + 0.44\zeff} + \frac{(0.52 + 0.086\sqrt{\nue})(1 + 0.87\ft)\nue}{1 + 1.13(\zeff - 1)^{0.5}},\\
  &\begin{aligned}\frac{\ft}{\fteff{32,\text{ee}}} = &1 + \frac{0.23(1 - 0.96\ft)\sqrt{\nue}}{\zeff^{0.5}} \\&+ \frac{0.13(1 - 0.38\ft)\nue}{\zeff^2}\left(\sqrt{1 + 2(\zeff-1)^{0.5}} + \ft^2\sqrt{\nue}\sqrt{0.075 + 0.25(\zeff - 1)^2}\right),
   \end{aligned}
  \\
  &\frac{\ft}{\fteff{32,\text{ei}}} = 1 + \frac{0.87(1 + 0.39\ft)\sqrt{\nue}}{1 + 2.95 (\zeff - 1)^2} + 1.53(1 - 0.37\ft)\nue (2 + 0.375[\zeff - 1]),\\
  &\alpha = \left(\frac{\alpha_0 + 0.7\zeff \ft^{0.5}\sqrt{\nui}}{1 + 0.18 \sqrt{\nui}} - 0.002(\nui)^2\ft^6\right) \frac{1}{1 + 0.004(\nui)^2\ft^6},\\
  &\alpha_0 = -\frac{0.62 + 0.055(\zeff - 1)}{0.53 + 0.17(\zeff - 1)} \times \frac{1 - \ft}{1 - (0.31 - 0.065[\zeff -1])\ft - 0.25\ft^2}.
\end{align}
\end{widetext}
The trapped particle fraction, denoted $f_{\text{trap}}$ in Ref.~\onlinecite{Redl}, is
\begin{equation}
\ft = 1 - \frac{3}{4} \left<B^2\right> \int_0^{1/B_{\max}} \frac{\lambda \, d\lambda}{\left<\sqrt{1-\lambda B}\right>}.
\end{equation}
The effective collisionalities are
\begin{align}
\nue&=6.921\times 10^{-18} \frac{n_e \zeff \ln\Lambda_e}{T_e^{3/2}}  \left(\frac{B_{\max} + B_{\min}}{B_{\max} - B_{\min}}\right)^{3/2} \frac{G + \iota I}{\iota - N} \lang \frac{1}{B}\rang, \\
\nui&=4.90\times 10^{-18} \frac{n_i \zeff^4 \ln\Lambda_{ii}}{T_i^{3/2} \epsilon^{3/2}} \left(\frac{B_{\max} + B_{\min}}{B_{\max} - B_{\min}}\right)^{3/2}\frac{G + \iota I}{\iota - N} \lang \frac{1}{B}\rang.
\end{align}
The Coulomb logarithms are \citep{sauter}
\begin{align}
    \ln\Lambda_{e} = & 31.3 - \ln{\left(\frac{\sqrt{n_e}}{T_e}\right)}\\   
  \ln\Lambda_{ii} = & 30 - \ln{\left(\frac{\zeff^3\sqrt{n_i}}{T_i^{3/2}}\right)},
\end{align}
with the temperatures in eV and the densities in m$^{-3}$.

In \eqref{eq:redl}, if light impurities are included in the calculation, the terms should be substituted as
\begin{align}
  p_{\text{i}}\dnidp &\to \sum_{\text{ions,impurities}} p_{a}\frac{\p \ln{n_{a}}}{\p \psi_{\text{t}}} \\
  p_{\text{i}}\dTidp &\to \sum_{\text{ions,impurities}} p_{a}\frac{\p \ln{T_{a}}}{\p \psi_{\text{t}}},
\end{align}
although this is only an approximation, as it assumes that the impurity and bulk ion transport coefficients (described by $\mcl_{31}$ and $\alpha$) are the same. This assumption is always somewhat violated in practice, if the impurities have different collisionalities than the bulk ions, but this can be expected to be partly compensated for by the fact that all the transport coefficients have been fitted to simulations with carbon impurities at different $\zeff$.


\bibliography{qs_stellarators_with_bootstrap}

\end{document}